\documentclass[conference]{IEEEtran}

\usepackage{url}
\usepackage{booktabs}   
\usepackage{subcaption} 

\usepackage{etoolbox}

\usepackage[utf8]{inputenc}
\usepackage[T1]{fontenc} 
\usepackage[english]{babel} 
\usepackage[inline]{enumitem}
\usepackage{xspace}
\usepackage{microtype} 
\usepackage{pifont}
\def\analyzed{\ding{88}}

\usepackage{tcolorbox} 

\usepackage{tikz}
\def\checkmark{\tikz\fill[scale=0.2](0,.35) -- (.25,0) -- (1,.7) -- (.25,.15) -- cycle;} 

\def\mysys{VulcaN\xspace}

\newcommand{\mypara}[1]{\vspace{5pt}\noindent{\textbf{#1}}}

\usepackage{graphicx}
\usepackage{xcolor,colortbl}
\usepackage{makecell} 
\usepackage{dashrule}
\usepackage[frozencache,cachedir=.]{minted}
\usepackage{inconsolata}

\usepackage{listings}
\usepackage{array}
\usepackage{tabularx}

\usepackage{booktabs}
\usepackage{arydshln}

\usepackage{multirow}
\usepackage{rotating}

\usepackage{ifthen}
\usepackage[noadjust]{cite}
\usepackage[numbers]{natbib}

\begin{document}
\title{Study of JavaScript Static Analysis Tools for Vulnerability Detection in Node.js Packages}

\author{Tiago Brito$^*$, Mafalda Ferreira, Miguel Monteiro, Pedro Lopes, Miguel Barros, José Fragoso Santos, Nuno Santos\\
\footnotesize{\{$^*$tiago.de.oliveira.brito, mafalda.baptista, miguel.figueiredo.monteiro, pedro.daniel.l, miguel.v.barros, jose.fragoso, nuno.m.santos\}@tecnico.ulisboa.pt}\\
\small{INESC-ID / IST, Universidade de Lisboa, Portugal}}

\maketitle

\thispagestyle{plain}
\pagestyle{plain}

\begin{abstract}
With the emergence of the Node.js ecosystem, JavaScript has become a widely-used programming language for implementing server-side web applications. In this paper, we present the first empirical study of static code analysis tools for detecting vulnerabilities in Node.js code. To conduct a comprehensive tool evaluation, we created the largest known curated dataset of Node.js code vulnerabilities. We characterized and annotated a set of 957 vulnerabilities by analyzing information contained in \textit{npm} advisory reports.
We tested nine different tools and found that many important vulnerabilities appearing in the OWASP Top-10 are not detected by any tool. The three best performing tools combined only detect up to 57.6\% of all vulnerabilities in the dataset, but at a very low precision of 0.11\%. Our curated dataset offers a new benchmark to help characterize existing Node.js code vulnerabilities and foster the development of better vulnerability detection tools for Node.js code.
\end{abstract}

\section{Introduction}

JavaScript has become one of the most popular programming languages for implementing server-side web applications. A driving factor in this trend has been the emergence of Node.js~\cite{nodejs}. Node.js is a cross-platform, back-end runtime environment that executes JavaScript code. Essentially, it can be used as a web container, housing JavaScript code that handles HTTP requests. Pivoted around Node.js, there is also an ecosystem of third-party packages managed by the Node Package Manager (\textit{npm}). Currently, \textit{npm} stores thousands of packages that web developers can readily import into their code, either for writing web applications or other packages.

The widespread adoption of Node.js makes the development of effective JavaScript vulnerability scanners a pressing matter. For one, the JavaScript~\cite{js} language features various constructs that display subtle behaviors. When employed by inexperienced code developers, these constructs may all too easily lead to the introduction of vulnerabilities. In addition, the manual detection of code vulnerabilities is complicated by the intricate \textit{npm} inter-package dependency system. In some cases, correct packages may become the source of security bugs as a result of ill-use by other packages. In others, buggy packages may end up propagating vulnerabilities up in the dependency chain to correct packages~\cite{zimmermann2019small}. This combination of factors opens up the path for serious security breaches in web applications. By exploiting security bugs, an attacker may be able to take over the entire server and/or affect many users through SQL injection, remote code execution, and other attacks~\cite{staicu2018synode,staicu2018freezing}.

An effective technique to prevent security vulnerabilities from creeping into production code is to integrate security analysis tools as part of Continuous Integration/Continuous Deployment (CI/CD) pipelines.
Using automatic vulnerability detection tools, developers can seamlessly receive prompt feedback about potentially existing security flaws in their code. This enables them to apply the necessary fixes at an early code development stage, thus helping them to improve the reliability of their software. In the same vein, JavaScript developers can benefit from code analysis tools that allow them to detect and fix security flaws inside \textit{npm} packages. Ideally, such tools should have high detection quality (i.e., low false-positive rate), and high coverage (i.e., low false-negative rate).

Motivated by this need, we set out to evaluate the effectiveness of existing JavaScript vulnerability detection tools at analyzing Node.js packages. We found a large body of work on client-side JavaScript security~\cite{stock2017web,lauinger2018thou,steffens2019don}, and some recent work in the study of vulnerabilities in \textit{npm} packages~\cite{staicu2018synode,staicu2018freezing}. However, no prior work has focused on evaluating tools that analyze server-side JavaScript code vulnerabilities, let alone on studying their effectiveness at finding security flaws in \textit{npm}  packages. As it turns out, performing this task is rather involved, given the absence of a gold standard for classifying such tools, and the lack of a comprehensive vulnerability dataset that can be used for benchmarking purposes.

In this paper, we present the first empirical study aimed at evaluating existing JavaScript vulnerability detection tools on Node.js packages. We focus exclusively on fully automatic, static code analysis tools that can be used in CI/CD pipelines. This excludes tools~\cite{staicu2018synode,gong:thesis:2018,gauthier2018affogato} that expect additional inputs, such as test suites, or tools that perform simple checks on known vulnerable dependencies~\cite{npm-audit,snyk}. In total, we screened 40 analysis tools for JavaScript and selected nine that can detect vulnerabilities at continuous integration time: NodeJsScan~\cite{njsscan}, CodeQL~\cite{codeql}, ODGen~\cite{odgen}, Graudit~\cite{graudit}, InsiderSec~\cite{insidersec}, ESLint SSC~\cite{eslint-ssc}, Microsoft's DevSkim~\cite{msdevskim}, Mosca~\cite{mosca} and Drek~\cite{drek}. We executed them against a curated dataset created by us containing \textit{npm} packages with annotated vulnerabilities, mainly: path traversal, cross-site scripting, insecure transfer using HTTP, resource exhaustion/denial-of-service, prototype pollution, OS command injection, code injection, and improper input validation. Then, we checked whether these tools can correctly identify these vulnerabilities.

Given that there is no curated dataset of Node.js vulnerabilities, our first step was to develop our own.
Building this dataset was in itself a challenging endeavor because we needed to identify real vulnerabilities in a large dataset of \textit{npm} packages. Our starting point was the \textit{npm} system itself. The \textit{npm} system runs a vulnerability report service that results in the generation of the so-called \textit{advisory} reports. These consist of textual descriptions of security vulnerabilities identified inside specific packages. These are real security vulnerabilities collectively identified by the Node.js developer community. Reports may also include an advice to upgrade the package to a fixed version. As such, advisory reports provide a reliable source for building our dataset. Unfortunately, these reports are not represented in a format that allows for automatic processing. Moreover, some of them may contain errors and therefore cannot be used unless a thorough analysis and verification are performed.

To overcome these difficulties, we manually analyzed 1359 advisory reports covering an equal number of vulnerable \textit{npm} package versions. These advisories represent 74\% of all the vulnerabilities officially reported inside benign \textit{npm} package versions until June 2021. In this process, we identified several anomalies in the advisory reports. We have then generated a curated dataset covering 957 of these advisories extended with annotations that specify the precise location of the reported code vulnerabilities. We found that the location of a large fraction of existing vulnerabilities can be fully expressed through \textit{source-sink} pair annotations. Our dataset can help the research community to i) characterize the vulnerabilities already detected within the \textit{npm} ecosystem, and ii) benchmark vulnerability detection tools. Our dataset is publicly available\footnote{\url{https://github.com/VulcaN-Study/Supplementary-Material}}.

We tested the pre-selected tools against our dataset and found they perform rather poorly, missing many vulnerabilities (low true positive rate/recall) and showing a high false positive rate (low precision). On average, they were able to correctly identify only 15.1\% of the total number of vulnerabilities in our dataset. The combination of the three best-performing tools detects 57.6\% of all vulnerabilities, albeit with only 0.11\% precision. The best performing tools, ESLint SSC and CodeQL, manage to detect 41.5\% and 31.3\% across all types of vulnerabilities and reach their peaks when it comes to identifying prototype pollution (79.2\% for CWE-471 and 86.1\% for CWE-1321) and path traversal (71.2\%) vulnerabilities, respectively. %
%
Of the 957 known vulnerabilities in the dataset, 324 (33.8\%) were not detected by any of the selected tools.
Some of the causes are tied to fundamental limitations of state-of-the-art code analysis techniques when it comes to analyzing server-side, JavaScript code vulnerabilities in the \textit{npm} ecosystem. Addressing these limitations is an interesting research direction for future work.

In summary, our paper makes four contributions: (i) a curated dataset with 957 real-world vulnerabilities in \textit{npm} package versions, which will be fundamental to evaluate future advancements in static analysis tools for detection of vulnerabilities in Node.js applications, (ii) a survey of existing vulnerability detection tools for JavaScript / Node.js code, (iii) a quantitative assessment of the vulnerability detection toolset against our curated dataset, and (iv) a study of the main causes of missing important vulnerabilities in \textit{npm} packages, which opens up several research avenues in this field.

\section{Study Design}

\subsection{Background}

Node.js features a package manager system named Node Package Manager (\textit{npm}), which currently stores thousands of third-party packages. Presently, it is difficult to guarantee the absence of security vulnerabilities in packages uploaded to \textit{npm} because there is no systematic code triage in place. Consequently, the \textit{npm} community mainly relies on third-party vulnerability reports to identify the potentially vulnerable packages that have already been included in the ecosystem.

\begin{figure}[]
    \centering
    \includegraphics[width=0.65\columnwidth]{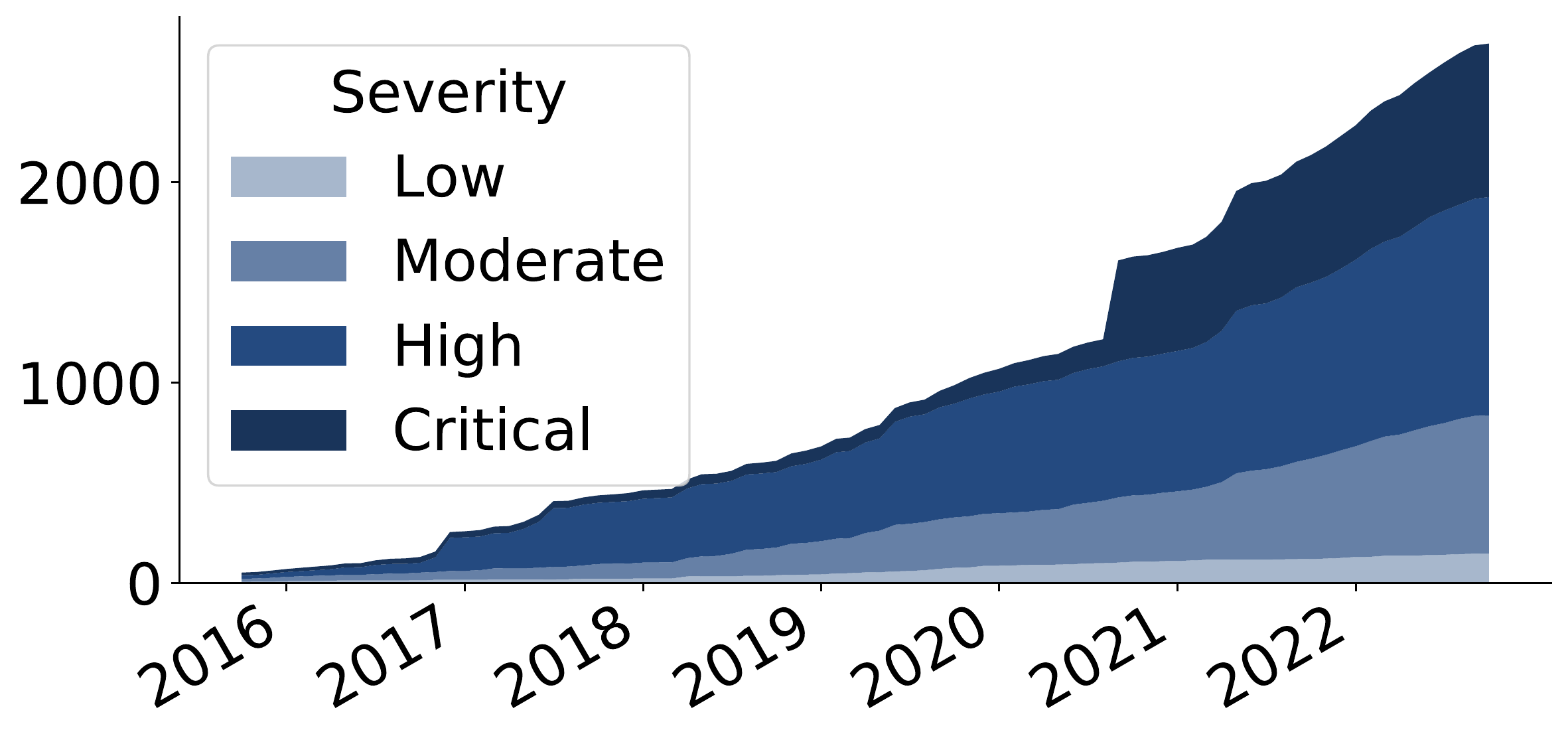}
    \caption{\small Evolution of published advisories over time.}
    \label{fig:advisories-over-time}
    \vspace{-0.3cm}
\end{figure}

The \textit{npm} system alerts JavaScript developers whenever they use a package version reported as vulnerable. These vulnerability alerts are called \textit{npm advisories}, as their purpose is to advise developers to update the vulnerable dependencies to either a fixed package version or to select another package entirely. When someone identifies a potential vulnerability in an \textit{npm} package, they produce a vulnerability report and submit it to the \textit{npm} security team, which checks the report, notifies package maintainers, and publishes the advisory, either when the package maintainers release a fix or if they remain unresponsive for longer than 45 days~\cite{npm-report-procedure}. Typically, an advisory report includes: package name, affected versions, description of the vulnerability, effects and references, commits, and/or code examples that help trigger the vulnerability. This information is then made available to developers in the advisory page (see example page for advisory 315 in~\cite{advisory-315}). Figure~\ref{fig:advisories-over-time} displays the evolution of the number of published advisories since \textit{npm}'s inception until October 11\textsuperscript{th} 2022, broken down according to their severity level. This number has steadily grown at nearly 40 advisories per month; about 70\% cover vulnerabilities considered to pose risks of high/critical severity.

\subsection{Research Questions and Scope}

In this work, we investigate four main research questions:

\mypara{RQ1. How to obtain an annotated dataset of vulnerabilities in server-side JavaScript code?} To evaluate JavaScript vulnerability detection tools, we require an annotated vulnerability dataset to compare the output of a given tool against ground truth data. The \textit{npm} repository provides an excellent source for retrieving both (i) an extensive collection of vulnerable Node.js applications (i.e., vulnerable \textit{npm} package versions), and (ii) information about real-world vulnerabilities (documented by the advisory database). Unfortunately, this information cannot be used as-is from existing advisories. First, advisories often lack relevant information about the reported vulnerability (e.g., the exact code location of the vulnerability within the package). Second, in many cases, most of the explanations regarding the reported vulnerability are given in external references, where information tends to be inconsistent and unstructured. Third, some advisories may be incorrect in places (e.g., the classification of the vulnerability type), which may lead to the mischaracterization of existing JavaScript vulnerabilities. These obstacles preclude an automated advisory analysis approach.

\mypara{RQ2. What is the state-of-the-art of existing security-orien-ted static analysis tools for Node.js code?} We are interested in assessing which static vulnerability detection tools are available. We need to distinguish between a broader set of code analysis tools which can serve many different purposes (e.g., detecting programming malpractices) from those that are specifically oriented toward the detection of vulnerabilities. Furthermore, we aim to analyze which code analysis techniques are employed by these tools. This will help us to better assess the strengths and weaknesses of each technique when evaluating each tool.

\mypara{RQ3. How effective are available detection tools in uncovering vulnerabilities in JavaScript code?} We are interested in empirically determining and characterizing how precise the publicly available vulnerability detection tools are at identifying vulnerabilities in known vulnerable JavaScript code.

\mypara{RQ4. What are the main reasons for missing the detection of vulnerabilities?} We aim to understand the key limitations of existing static vulnerability detection tools that explain their failure to detect known vulnerabilities in JavaScript code.

\vspace{2pt} The research questions above have a twofold goal: (i) characterize vulnerabilities in \textit{npm} packages in the wild, and (ii) evaluate the effectiveness of existing static JavaScript vulnerability detection tools. To better set the reader's expectations about our study, 
we further clarify these subgoals and discuss other relevant directions we left outside the scope of this work.

Firstly, our study is narrowed toward the characterization of vulnerabilities reported inside \textit{npm} packages. As such, our results cannot be extrapolated to characterize the typical programming flaws introduced by JavaScript developers during the code development stage; nor is this our goal. Developers may use vulnerability detection tools in their continuous integration frameworks that may capture some vulnerabilities before the code is deployed to \textit{npm}. 
This means that some security flaws may have been fixed prior to deploying the code into production. Also note that, given that we only analyze formerly identified vulnerabilities, our curated dataset may not be representative of all existing JavaScript vulnerabilities lingering inside \textit{npm} packages; this is also not our purpose. In contrast, we intend that our curated dataset contains a large set of confirmed real-world vulnerabilities that can be used for assessing existing (and future) vulnerability detection tools.

Secondly, we focus exclusively on \textit{fully automatic vulnerability analysis tools} that can be easily integrated into existing code review pipelines. This excludes the only three existing dynamic analysis tools for detecting vulnerabilities in Node.js code: SyNode~\cite{staicu2018synode}, NodeSec\cite{gong:thesis:2018} and Affogato~\cite{gauthier2018affogato}. These tools require a set of unit tests covering all security sensitive behaviors of the package to be analysed. Such test suites must be manually written as the automatic generation of high-coverage test suits for Node.js applications is still an open problem. Nevertheless, our curated dataset can still be used as a benchmark for evaluating the effectiveness of these excluded tools.

\subsection{Study Methodology}

Our approach to answering the questions above is to perform an empirical study consisting of the following three tasks:

\mypara{Task 1. Manual analysis of advisory reports and building the annotated dataset:} We manually analyzed all advisories, filling in the missing information, namely by identifying the code location that triggers the vulnerability. This information is part of our curated dataset used in the following tasks.

\mypara{Task 2. Selection and execution of code analysis tools for vulnerability detection in JavaScript code:} We searched both the literature and the web for code analysis tools that can be used for vulnerability detection. We executed all of the selected tools on all the vulnerable packages in our curated dataset.

\mypara{Task 3. Evaluation of tool output according to the ground truth extracted from the annotated dataset:} We automatically compare a tool's result with the corresponding dataset annotations, considering the verbosity levels of each tool.

\begin{figure}[t]
    \centering
    \includegraphics[width=0.78\columnwidth]{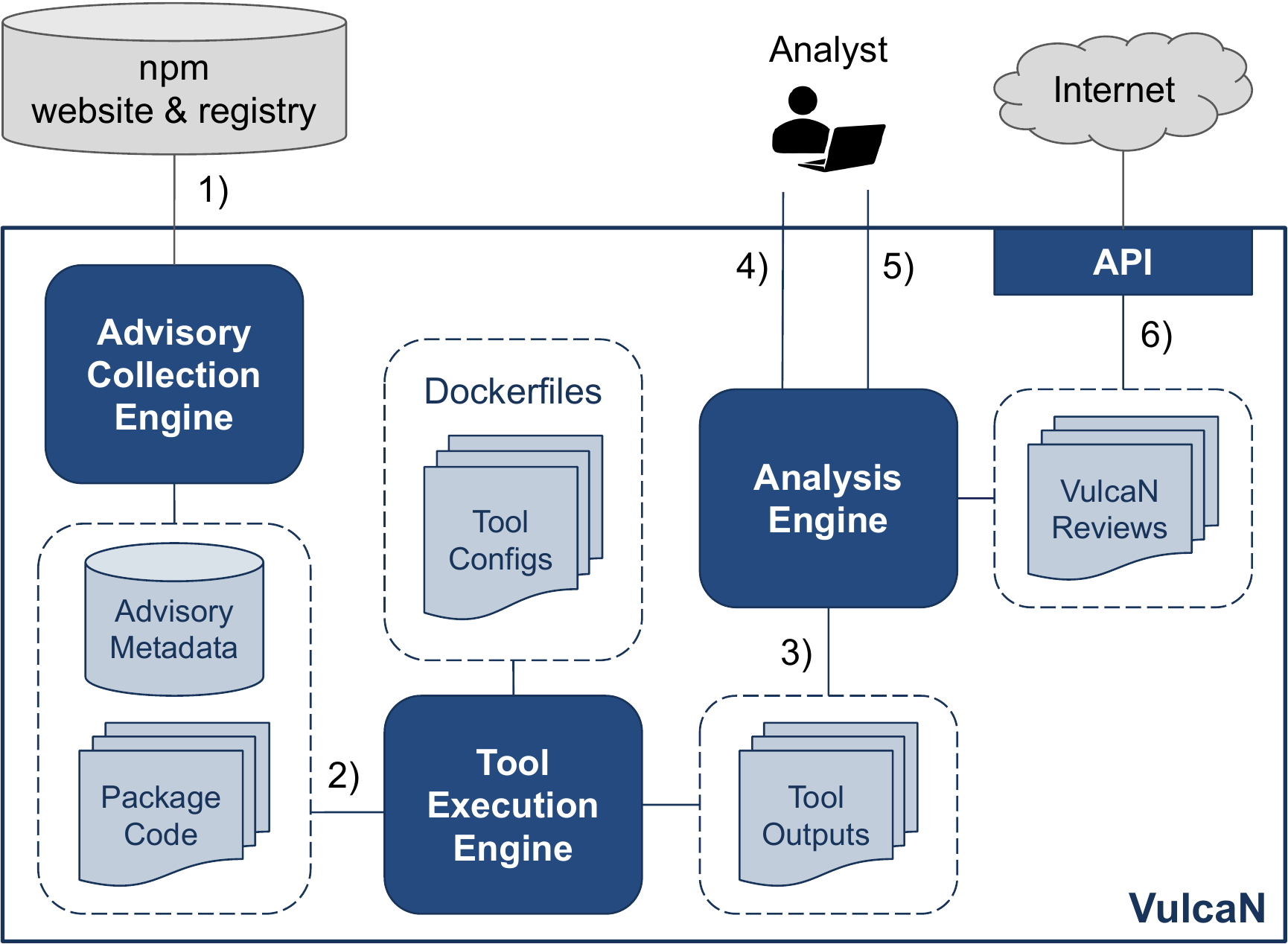}
    \caption{\small \mysys architecture.}
    \label{fig:vulcan}
        \vspace{-0.2cm}
\end{figure}

\vspace{2pt} To support our methodology, we developed \mysys, a testbed for analyzing \underline{vul}nerability detection tools for \underline{N}ode.js packages. \mysys is an execution and analysis framework to collect vulnerable versions of \textit{npm} packages, run vulnerability detection tools over all collected packages, and help security analysts perform the vulnerability analysis. Currently, \mysys supports nine tools (see Section~\ref{section:tools}) and has these main features:

\begin{itemize}
    \item an interface to download the latest published advisories;
    \item a Docker-based extension for adding more analysis tools;
    \item an interface for accessing both the metadata and the output of the analysis tools for each advisory; 
    \item an interface for annotating each advisory with a review file, which contains the code location of the vulnerability and the classification of the results of the analysis tools;
    \item an interface for automatically comparing the output of the tools with the corresponding review in our curated dataset, this includes a parser for the ouput of each tool;
    \item a web API exposing the curated dataset and the classification of the analysis tools.
\end{itemize}

Figure~\ref{fig:vulcan} shows the architecture of \mysys and how each component is used in the context of our empirical study. First, the \textit{Advisory Collection Engine} crawls the \textit{npm} advisory website~\cite{npm-advisories} and collects the available metadata about all published advisories saving it in a \textit{MongoDB} database (1). The latest vulnerable version for each advisory, referenced in the collected metadata, is then downloaded via the \textit{npm} registry. Second, the \textit{Tool Execution Engine} builds a docker image for each tool, according to the specified \textit{Dockerfile}, and runs the respective container for all downloaded packages, storing the output of each tool locally (2). Third, the advisory metadata, code, and tool outputs are fed to the \textit{Analysis Engine}, which generates an environment for advisory report analysis (3). The analyst accesses the environment (4) and produces a review file comprising an accurate description of the vulnerable code location, which is then submitted to the framework (5). Finally, the \textit{Analysis Engine} processes the submitted information, automatically compares the tools' outputs with the analyst reviews (dataset) using a dedicated parser for each tool, and exposes it through a web API (6).

\section{Dataset of Vulnerabilities (RQ1)}
\label{section:dataset}

In this section, we address RQ1, explaining how we created our curated dataset of JavaScript code vulnerabilities.

\subsection{Selection and Validation of Reports}

To create our dataset, we collected a snapshot of the existing \textit{npm} advisories until the end of June 2021. Then, through manual analysis, we excluded some advisories and fixed inconsistencies in the remaining ones (see Table~\ref{tab:anomalies}). Out of the 1828 advisories from the original snapshot, we excluded 469, keeping 1359 for further analysis. Next, we present our exclusion criteria and discuss the detected inconsistencies.

\mypara{Excluded advisories.} As of June 30\textsuperscript{th} 2021, there were 1828 advisories published in \textit{npm}.
Of these 1828 advisories, 416 are categorized by \textit{npm} as Embedded Malicious Code (CWE-506): these are packages designed with malicious intent, named very similarly to real legitimate packages so as to deceive developers into installing them. These packages are not relevant for our study, which focuses only on unintentional vulnerabilities. From the remaining 1412 advisories, we excluded 31 for lacking available code. Lastly, out of the resulting 1381 vulnerable package versions, 22 were excluded for not including JavaScript code; instead, they had pre-transpiled variants such as CoffeeScript~\cite{coffeescript} and TypeScript~\cite{typescript:toot:2016}, which prevented us from analyzing their source code directly. Consequently, in the end, \mysys successfully collected 1359 advisories and their corresponding package versions for further manual analysis.

\begin{table}[]
\centering
\small
\begin{tabular}{lc}
\hline
\textbf{Exclusions \& Inconsistencies}           & \textbf{\# of Advisories} \\ \hline
Malware Packages             & 416                 \\
Missing Package Code         & 31                  \\
Missing JavaScript Code      & 22                  \\ \hline
Incorrect Vulnerable Version & 42                 \\
Missing External References  & 291                 \\ 
Imprecise CWE                & 101                  \\ \hline
Lack of Analysis Information & 402  \\ \hline
\end{tabular}
\caption{\small Number of advisories excluded or inconsistent.}
\label{tab:anomalies}
\vspace{-0.3cm}
\end{table}

\mypara{Detected inconsistencies.} During the manual analysis, we noticed several inconsistencies in the collected advisories. Most notably, only a small minority of advisories comes with the exact code locations that trigger their corresponding vulnerability. Furthermore, some advisories provide an \textit{incorrect vulnerable package version}, i.e., the advisory metadata points to a package version that does not contain the described vulnerability. When the advisory does not come with additional external references, which is the case for 21\% of the analyzed advisories, correcting the incorrect vulnerable package version anomaly can be quite challenging, as the advisory metadata alone is generally insufficient for pinpointing the correct package version. Another detected anomaly is the \emph{imprecise classification of vulnerability type/category}. Most of the times this imprecision is subjective, as a Common Weakness Enumeration (CWE)~\cite{cwe} class can be a subcategory (child) of another more general CWE. This is particularly common for Path Traversals (CWE-22) and Code Injections (CWE-94), to which more precise classes can be attributed; in particular, CWE-23 (Relative Path Traversal) and CWE-24 (another specific Path Traversal variant) to CWE-22 and CWE-95 (Eval Injection) to CWE-94. Some vulnerabilities are simply miscategorised; for instance, sometimes Code Injection (CWE-94) vulnerabilities are categorized as Cross-Site Scripting (CWE-79). During the analysis, we detected 63 cases of vulnerability miscategorization (different CWE) and 21 cases of incorrect vulnerable package version referenced in the advisory. The remaining cases lack the CWE categorization.

\subsection{Analysis of Reported Vulnerabilities}

We analyzed the vulnerabilities in the selected 1359 \textit{npm} advisories. Our goal was to characterize the vulnerability landscape of the \textit{npm} package ecosystem by studying the distribution of existing vulnerabilities according to their category and assess the potential security risks posed by the affected packages. From the 1359 advisories manually analyzed, we managed to verify the vulnerability for 957 advisories: these are the ones included in our dataset and characterized in this study. The remaining advisories (402) did not include sufficient information to successfully verify the vulnerability.

\begin{figure}[t]
    \centering
    \includegraphics[width=0.75\columnwidth]{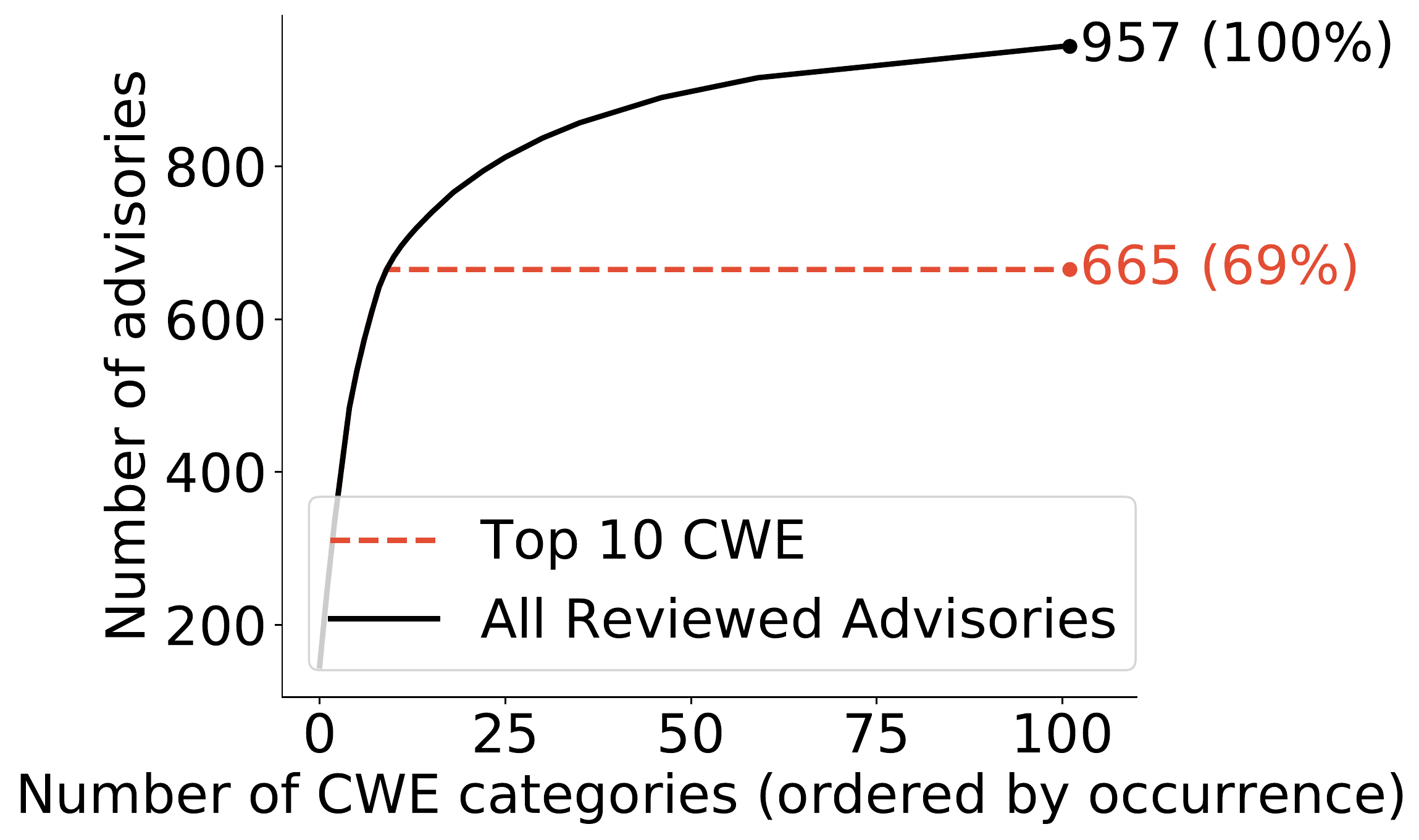}
    \caption{\small CDF of \# of reviewed advisories ranked by CWE.}
    \label{fig:cwe-cdf}
        \vspace{-0.1cm}
\end{figure}

Figure~\ref{fig:cwe-cdf} displays the cumulative distribution function (CDF) of the number of vulnerabilities of our dataset ranked by their CWE category. This distribution is heavily skewed toward a relatively small number of CWE categories, i.e., a large fraction of vulnerabilities pertains to a restricted set of categories. In particular, the top-10 CWEs cover 665 advisories, i.e., 69\% of the total number of verified vulnerabilities.

To estimate the potential security risks of such vulnerabilities, we mapped each of the top-10 CWE categories to the latest OWASP ranking (from 2021). OWASP is an organization that works to raise awareness about web security and ultimately improve it. The OWASP Top 10~\cite{owasp-top10} list is a popular document representing a broad consensus about the most critical security risks to web applications. Updated every few years, this document describes risks, such as injection attacks, broken authentication, and known vulnerable dependencies. As shown in Table~\ref{table:cwes}, most vulnerability types can be mapped to a top-10 web security risk. This means that \textit{npm} packages have well-known risks that security professionals are familiar with.
Most notably, 9 out of 10 CWE categories (i.e., 576 advisories) appear in the top-3 OWASP list. This translates to approximately 60\% of the total number of analyzed vulnerabilities in our dataset (957). In other words, many reported vulnerabilities can introduce serious security flaws in web applications.

\subsection{Our Curated Dataset}

Based on the selected advisories, we created a curated dataset aimed at providing a baseline for assessing the effectiveness of vulnerability detection tools. It comprises: (i) the code for the vulnerable version of the \textit{npm} package indicated in the advisory, and (ii) a corresponding \textit{review} file. This file contains ground truth information that allows us to validate the output of a given tool when analyzing that specific package version.

\mypara{Review example:} Listing~\ref{snippet:review} shows the created review file for advisory 315~\cite{advisory-315}. This advisory reports the presence of a code injection vulnerability in package \textit{summit-0.1.22}.
A review is a JSON object that contains several fields that describe: i) the advisory identifier (\texttt{id}), ii) the vulnerability type as per the CWE taxonomy (\texttt{cwe}), iii) the affected package version (\texttt{package\_link}), and iv) the vulnerability location expressed as a \texttt{source}/\texttt{sink} pair. A source/sink is specified as a JSON object with fields denoting: the file name (\texttt{file}); the line number (\texttt{lineno}); and the corresponding line of code (\texttt{code}).

\begin{table}[t]
\centering
\small
\begin{tabular}{c|llc}
\hline
\textbf{\#} & \textbf{CWE} & \textbf{Security Risk}  & \textbf{\# Occurrences} \\ \hline
1 & CWE-22  & 1. Broken Access Control      & 146   \\ 
2 & CWE-79  & 3. Injection                  & 99   \\ 
3 & CWE-400 & -                             & 89    \\ 
4 & CWE-78  & 3. Injection                  & 75    \\ 
5 & CWE-818 & 2. Cryptographic Failures     & 75   \\ 
6 & CWE-471 & 3. Injection                  & 48    \\ 
7 & CWE-20  & 3. Injection                  & 41    \\ 
8 & CWE-1321 & 3. Injection                 & 36     \\
9 & CWE-94  & 3. Injection                  & 33    \\ 
10 & CWE-77  & 3. Injection                 & 23     \\\hline 
\end{tabular}
\caption{\small Possible mapping of most occurring vulnerabilities in dataset with OWASP Top 10 Web Security Risks (2021)~\cite{owasp-mapping}: Path Traversal (CWE-22), Cross-site Scripting (CWE-79), Resource Exhaustion (CWE-400), Insufficient Transport Layer Protection (CWE-818), OS Command Injection (CWE-78), Modification of Assumed-Immutable Data (CWE-471), Improper Input Validation (CWE-20), Improperly Controlled Modification of Object Prototype Attributes (CWE-1321), Code Injection (CWE-94), and Improper Neutralization of Special Elements used in a Command (CWE-77).}
\label{table:cwes}
\vspace{-0.1cm}
\end{table} 

\begin{listing}[t]
\begin{minted}[
    breaklines,
    breakbytoken,
    fontsize=\footnotesize]{json}
{
  "advisory": { "id": 315, "cwe": "CWE-94" },
  "package_link": "registry.npmjs.org/summit/-/summit-0.1.22.tgz",
  "vulnerability": [
    {
      "source": {
        "file": "lib/drivers/search/pouch.js",
        "lineno": 4,
        "code": "return function search (opts) {"
      },
      "sink": {
        "file": "lib/drivers/search/pouch.js",
        "lineno": 20,
        "code": "eval(opts.filter);"
      }
    }
  ]
}
\end{minted}
\caption{\small Example of \mysys review file for advisory 315.}
\label{snippet:review}
\vspace{-0.5cm}
\end{listing}

\mypara{Vulnerability location:} The location fields are used to determine if the output of a given vulnerability detection tool is correct. Besides using (one or multiple) \texttt{source}/\texttt{sink} pairs to locate a vulnerability, we can also employ (one or multiple) \texttt{block} patterns, which indicate contiguous code regions in which the flaw exists. This form of representation is necessary for vulnerability types that cannot be expressed as source/sink pairs, e.g., usage of HTTP instead of HTTPS which allows for MITM attacks (CWE-818).
Interestingly, we found that, in most cases (78\%), the exact location of a vulnerability is very clear, e.g., a call to \texttt{eval} in a code injection. These cases can be represented by source/sink pairs, while the remaining cases (22\%) can be represented using code blocks that cover all vulnerability-relevant code. Although the block size depends on the vulnerability, in our dataset the average block size is six lines-of-code.
Moreover, since the review files can be processed automatically, we believe that our curated dataset will be useful for benchmarking purposes beyond the scope of our work.

\begin{table*}[t]
\centering
\footnotesize
\resizebox{\textwidth}{!}{
\begin{tabular}{l||ccccc}
\hline
    \textbf{Included Tools} &
    \textbf{\makecell[l]{Only Package\\Source Code (C0)}} &
    \textbf{\makecell[l]{Not Available\\or Proprietary (C1)}} &
    \textbf{\makecell[l]{Not Scriptable\\Interface (C2)}} &
    \textbf{\makecell[l]{Not Security\\Oriented (C3)}} &
    \textbf{\makecell[l]{Other Exclusionary\\Reasons}}\\ \hline
    \textbf{\begin{tabular}[c]{@{}l@{}}
        NodeJsScan~\cite{njsscan} \analyzed\\
        CodeQL~\cite{codeql} \analyzed\\
        ODGen~\cite{odgen} \analyzed\\
        Graudit~\cite{graudit} \analyzed\\
        InsiderSec~\cite{insidersec} \analyzed\\
        ESLint SSC~\cite{eslint-ssc} \analyzed\\
        MS DevSkim~\cite{msdevskim} \analyzed\\
        Mosca~\cite{mosca} \analyzed\\
        Drek~\cite{drek} \analyzed
    \end{tabular}} &
    \begin{tabular}[c]{@{}l@{}}
        SyNode~\cite{staicu2018synode}\\
        NodeSec~\cite{gong:thesis:2018}\\
        Affogato~\cite{gauthier2018affogato}\\
    \end{tabular} &
    \begin{tabular}[c]{@{}l@{}}
        Beyond Security~\cite{beyond-security}\\
        Checkmarx~\cite{checkmarx}\\
        Fortify~\cite{fortify}\\
        Veracode~\cite{veracode}\\
        Kiuwan~\cite{kiuwan}\\
        CodeSonar~\cite{codesonar}\\
        Thunderscan~\cite{thunderscan}\\
        WhiteHat~\cite{whitehat}
    \end{tabular} &
    \begin{tabular}[c]{@{}l@{}}
        JsPrime~\cite{jsprime} \analyzed\\
        Codeburner~\cite{codeburner} \analyzed\\
        SonarQube~\cite{sonarqube} \analyzed\\
        CodeWarrior~\cite{codewarrior} \analyzed
    \end{tabular} &
    \begin{tabular}[c]{@{}l@{}}
        WALA~\cite{wala}\\
        PMD~\cite{pmd}\\
        Aether~\cite{aether}\\
        Coala~\cite{coala}\\
        JsHint~\cite{jshint} \analyzed\\
        EsComplex~\cite{escomplex}\\
        Coverty Scan~\cite{coverty-scan}\\
        DeepScan~\cite{deepscan}\\
        AppInspector~\cite{appinspector} \analyzed\\
        TAJS~\cite{tajs2009}\\
        SAFE~\cite{lee2012safe}\\
    \end{tabular} &
    \begin{tabular}[c]{@{}l@{}}
        ESLint SP~\cite{ESLint-security} \analyzed\\
        Mozilla ScanJs~\cite{mozilla-scanjs} \analyzed\\
        SemGrep~\cite{semgrep}\\
        JAW~\cite{jaw} \analyzed\\
        Joern~\cite{yamaguchi2014modeling,joern-repo} \analyzed
    \end{tabular} \\ \hline
\end{tabular}
}
\caption{\small Tools included in / excluded from this study. Excluded for reasons beyond C0, C1, C2, \& C3: \textit{ESLint Security Plugin} and \textit{Mozilla ScanJs} use \textit{ESLint} rules subsumed by \textit{ESLint SSC}'s; \textit{SemGrep} requires specially-crafted rules for security purposes and is the backbone of NodeJsScan; JAW only implements rules for detecting client-side CSRF; Joern supports JavaScript inspection, but does not support default rules for detection. Some tools excluded due to C1 were tested using free trials, but failed to comply with additional criteria.}
\label{table:all-tools}
\vspace{-0.3cm}
\end{table*}

\mypara{Methodology:} Vulnerability locations were identified manually. To account for their possible mislabeling, each advisory was analyzed by two authors at separate times and their results cross-checked. Two authors performing cross-validation disagreed in 84 reviews (8.8\% of 957 reviews). Most inconsistencies were differences in source/sink pairs. These cases were resolved by selecting the correct source/sink pair or specifying a superset of source/sink pairs. In rare cases, one author failed to locate the vulnerability. These cases were handled by jointly reviewing the location identified by the other author.

\mypara{Dataset size:} From the 1359 analyzed advisories, we were able to manually verify 957 review files (70\%) at the time of this paper submission. For each review file we confirmed the exact location of the reported vulnerability. For this reason, we are confident to include these reviews in our curated dataset.

\section{Vulnerability Detection Tools (RQ2)}
\label{section:tools}

We now focus on RQ2, explaining how we selected the tools considered in our study. We specify our eligibility criteria, survey the existing tools that satisfy them, and classify these tools according to the detection technique that they employ.

\subsection{Tool Selection Criteria and Selection Process}

We focus specifically on fully automatic tools for analysis of \textit{npm} packages. In particular, to select a given tool, it must:

\mypara{C0. Depend only on the package source code:} The tool requires only the source code of the package to analyze. This excludes tools that require a test suite to guide the analysis.

\mypara{C1. Be available and transparent:} The tool is publicly available and implements a technique that is non-proprietary. Its source code does not need to be open as long as the tool's code analysis techniques can be clearly characterized, e.g., through available documentation, rule set, and usage examples.

\mypara{C2. Have a scriptable interface:} The tool must support a command-line interface (CLI), or similar interaction, allowing it to be executed and its output analyzed via a script. This facilitates the scalability and automation of the analysis.

\mypara{C3. Be security-oriented:} The tool must identify vulnerabilities or security bad practices in JavaScript. This excludes tools that only construct artifacts, such as control-flow graphs, produce warnings about coding styles and conventions, or produce statistical information about the code, such as code metrics, that might be irrelevant from a security standpoint.

\vspace{2pt} Based on the above criteria, we ended up selecting nine tools for our testing purposes. We started by examining the academic literature~\cite{gong:thesis:2018,staicu2018synode,gauthier2018affogato,jaw,yamaguchi2014modeling,odgen} and searching the Internet, including OWASP lists~\cite{owasp-scanning-tools,owasp-free-opensource-tools}, repository collections~\cite{awesome-list-1,awesome-list-2,awesome-list-3} and other websites~\cite{tools-website-1,tools-website-2,tools-website-3}, for suitable tools for vulnerability detection in server-side JavaScript code. Most tools we screened were developed by the industry and the open-source community. In total, we first collected 40 JavaScript analysis tools. This full list is presented in Table~\ref{table:all-tools}.

After inspecting all 40 analysis tools, we found that we needed to manually test 19 tools, which are annotated with the symbol \analyzed~in Table~\ref{table:all-tools}. These 19 tools were tested against the Damn Vulnerable Node Application (DVNA)~\cite{dvna}, a web application written in JavaScript that was purposely built with a range of vulnerabilities matching the OWASP Top 10 Web Security Risks. After executing each of the 19 tools against the vulnerable application, we excluded those that do not allow the analysis to be automated via a script and also those that fail to show security-oriented results.

Out of the remaining 19 tools, we selected 14 tools that are available, transparent, can be automated, and show security-oriented results. Out of these 14 tools, we also excluded \textit{ESLint Security Plugin}, \textit{Mozilla ScanJs}, \textit{SemGrep}, \textit{JAW}, and \textit{Joern} as explained in the caption of Table~\ref{table:all-tools}. Consequently, we ended up with 9 distinct candidates that represent proper fully automatic vulnerability detection tools for Node.js applications.

\subsection{Detection Techniques}
\label{sec:methodologies}

By manually analyzing the source code of the nine selected tools, we characterized them according to the employed technique for finding vulnerabilities. It is important to understand how these techniques work as they can have a significant impact on the effectiveness of the tools. We categorize these techniques using three classes: \textit{graph-based analysis}, \textit{syntax-based analysis} and \textit{keyword-based analysis}.
Table~\ref{table:tool-method} maps every selected tool to its corresponding detection technique.

\begin{table}[t]
\centering
\small
\begin{tabular}{cll}
\hline
\textbf{Technique} &
  \textbf{Tool} &
  \textbf{Version} \\ \hline
\multirow{2}{*}{\makecell{Graph-based\\Analysis}} &
  CodeQL & 2.2.6\\
 & 
  ODGen & --\\ \hline
\multirow{2}{*}{\makecell{Syntax-based\\Analysis}} &
  NodeJsScan & 0.2.8 \\ 
 &
  ESLint SSC & ** \\ \hline
\multirow{5}{*}{\makecell{Keyword-based\\Analysis}} &
  Graudit & 2.8 \\ 
 &
  InsiderSec & 2.0.5 \\ 
 &
  MS DevSkim & 0.4.109 \\ 
 &
  Mosca & 0.8 \\ 
 &
  Drek & 1.0.3 \\ \hline
\end{tabular}
\caption{\small Vulnerability detection technique employed by each tool. **ESLint SSC was used with eslint@7.32.0.}
\label{table:tool-method}
\vspace{-0.5cm}
\end{table}

\mypara{Graph-based analysis tools} work by first constructing a graph-based model of the program to be analyzed. Such models usually coalesce into a single graph-like data structure various types of statically computed program artefacts, including: abstract syntax trees, control-flow graphs, and dependency graphs. The obtained data structure can be then inspected using queries written in domain-specific languages (DSL) especially designed for specifying vulnerable code patterns. For instance, typical queries aim at identifying code-flow paths through which user-controllable inputs can reach dangerous sinks. CodeQL is one of such tools that models source code as database records. These records can be queried using SQL-like statements that are specified in the form of rules/queries.

\mypara{Syntax-based analysis tools} employ a technique that searches the code to be analyzed for insecure syntax-aware patterns. Patterns can express simple control-flow conditions, e.g., calling a function with a particular variable. In contrast to graph-based analysis, this technique operates directly on source code and does not typically cater for more intricate dependency analysis or for matching patterns across multiple files. NodeJsScan is an example of such tools. It is used in GitLab's CI/CD~\cite{gitlab-cicd}.

\mypara{Keyword-based analysis tools} employ a code analysis technique that searches the code to be analyzed for strings associated with potentially insecure code. This search is typically performed through the use of regular expressions. Note that keyword-based analysis does not model the AST of the program to analyze. Consequently, it is considerably less expressive than graph- and syntax-based analysis, as it cannot reason about fine-grained control-flow interactions, often operating on a single line of code at a time.

\subsection{How Different Detection Techniques Work}
\label{sec:detection-techniques-explained}

To better understand how the aforementioned vulnerability detection techniques work, we examine, as an example, the advisory 315 of our dataset. According to the information available in the advisory page~\cite{advisory-315}, this example consists of a code injection vulnerability (CWE-94), which allows a malicious user to run arbitrary code on the targeted execution platform. In this type of attack the adversary is only limited by the expressiveness of the injected language. JavaScript code injections at the server can have more significant impact than those at the client-side, given that Node.js has fewer security barriers (e.g., no sandbox), and a larger and privileged API facilitating access to critical system resources (e.g., file system).

\begin{listing}[t]
\begin{minted}[
    xleftmargin=15pt,
    breaklines,
    breakbytoken,
    linenos,
    fontsize=\footnotesize]{js}
function search (opts) {
  if (!opts.filter && opts.collection) {
    if (typeof opts.collection === 'string') {
      opts.filter = "function filter (doc) { return doc.type === '" + opts.collection + "'}";
    } else { ... }
    eval(opts.filter);
    opts.filter = filter;
  }
}
\end{minted}
\caption{\small Code injection vulnerability (\textit{npm} advisory 315).}
\label{snippet:advisory-315}
\vspace{-0.2cm}
\end{listing}
\begin{listing}[t]
\begin{minted}[
    xleftmargin=15pt,
    breaklines,
    breakbytoken,
    linenos,
    fontsize=\footnotesize]{js}
opts.collection = `'};
const exec = require("child_process").exec;
exec("cat /etc/passwd", (err, stdout, stderr) => {
console.log(stdout); }); var a={ hello: 'world`;
search(opts)
\end{minted}
\caption{\small Exploit for code injection (\textit{npm} advisory 315).}
\label{snippet:poc-315}
\vspace{-0.2cm}
\end{listing}

The vulnerable code snippet of the advisory 315 is given in Listing~\ref{snippet:advisory-315}. In this example, the variable \textit{opts} is bound to a user-controlled object with the properties \textit{filter} and \textit{collection}, which can be trivially tainted with a maliciously crafted input to produce valid JavaScript code that reaches the \textit{eval} function. One can leverage this vulnerability to execute arbitrary JavaScript code, including OS-level commands by using a payload like the one given in Listing~\ref{snippet:poc-315}.

\mypara{Using graph-based analysis:}
Both CodeQL and ODGen adopt graph-based analysis. Listing~\ref{rule:codeql-315} shows an example of a CodeQL rule designed to detect calls to the \textit{eval} function using user-controlled inputs. In order to create this CodeQL rule, one starts by specifying the appropriate configuration, that is, a code description of the targeted sources and sinks. In this case, we are interested in code flows from \emph{remote flow sources}, described by the predicate \emph{isSource} (lines 3 to 5), to the \textit{eval} sink, described by the predicate \emph{isSink} (lines 6 to 8). Then the main query (lines 11 to 13) states that, using the specified configuration (\textit{EvalTaint cfg}), CodeQL should find code paths from the specified source to the specified sink. The output of this query is a string with a description of the source-sink pairs that match the query. This particular rule is a simplified version of one of the CodeQL rules~\cite{codeql-rules-repo} executed inside \mysys.

\begin{listing}[t]
\begin{minted}[
    xleftmargin=15pt,
    breaklines,
    breakbytoken,
    linenos,
    fontsize=\footnotesize]{js}
class EvalTaint extends TaintTracking::Configuration {
  EvalTaint() { this = "EvalTaint" }
  override predicate isSource(Node node) {
    node instanceof RemoteFlowSource
  }
  override predicate isSink(Node node) {
    node = globalVarRef("eval").getACall().getArgument(0)
  }
}

from EvalTaint cfg, Node source, Node sink
where cfg.hasFlow(source, sink)
select sink, "Eval with user input from \$@.", source
\end{minted}
\caption{\small CodeQL rule for \textit{eval} taint-tracking~\cite{codeql-eval-taint-example}.}
\label{rule:codeql-315}
\vspace{-0.2cm}
\end{listing}

In general, graph-based analysis works well for taint-tracking, but it requires every source and sink to be explicitly encoded into rules. These sources and sinks change over time as languages evolve and new popular third-party packages are created. This is why the community has started to work on automatically generating such taint-tracking specifications~\cite{staicu2020extracting}.

\begin{listing}[t]
\begin{minted}[
    xleftmargin=15pt,
    breaklines,
    breakbytoken,
    linenos,
    fontsize=\footnotesize]{yaml}
patterns:
  - pattern-inside: function $FUNC($REQ, $RES, ...) {...}
  - pattern-either:
      - pattern: eval(..., <... $REQ.$QUERY ...>, ...)
\end{minted}
\caption{\small NodeJsScan rule for \textit{eval} detection~\cite{njsscan-eval-taint-example}.}
\label{rule:njsscan-315}
\vspace{-0.2cm}
\end{listing}

\mypara{Using syntax-based analysis:}
NodeJsScan helps us showcase syntax-based analysis. Listing~\ref{rule:njsscan-315} lists an excerpt of the NodeJsScan rule that detects potentially vulnerable uses of \textit{eval}. To be applied, this rule must match two related patterns. First, \textit{eval} must occur inside a function receiving two or more arguments (line 2). Then, \textit{eval} must be called with a parameter computed using one of the given arguments (line 4). NodeJsScan includes analogous rules for other vulnerability types~\cite{njsscan-rules-repo}.

Similarly to the previous technique (graph-based), syntax-based analysis also suffers from the source-sink specification limitation. Additionally, there are some other limitations specific to syntax-based analysis. For example, it may lead to a high number of false positives, as it is not expressive enough to capture the dependencies of the variables occurring in the patterns; e.g., it will detect all calls to \textit{eval} regardless of whether or not their given input can be controlled by the user. Furthermore, it commonly leads to \textit{rule overfitting}, resulting in over-specific rules that match known examples of vulnerabilities, but are not general enough to capture other instances of the same vulnerability. 
In Listing~\ref{rule:njsscan-315}, the \textit{eval} call is only detected when it occurs inside the body of a specific type of function  declaration. Besides ignoring \textit{eval} calls at the top level, this pattern also ignores calls to \textit{eval} which occur inside the body of JavaScript functions declared using alternative syntactic constructs (e.g. function constructor and lambdas).

\mypara{Using keyword-based analysis:}
The tool we use to illustrate keyword-based analysis is Graudit. The following is an excerpt of a Graudit rule that detects the use of \textit{eval}:

\begin{minted}[
    xleftmargin=15pt,
    breaklines,
    breakbytoken,
    fontsize=\footnotesize]{text}
eval[[:space:]]*\(
\end{minted}

Here, we see a regular expression pattern that simply detects any call to the \textit{eval} function. This technique suffers from several limitations such as the source-sink specification problem and a high number of false positives. Graudit includes many other rules for dangerous sinks in Node.js applications~\cite{graudit-rules-repo}.

\begin{listing}[t]
\begin{minted}[
    %xleftmargin=15pt,
    breaklines,
    breakbytoken,
    fontsize=\footnotesize]{XML}
<report_mosca>
    <Path>/src/lib/drivers/search/pouch.js</Path>
    <Title>Possible code injection</Title>
    <Description>
        Command injection is an attack in which the goal is execution of arbitrary commands on the host operating system via a vulnerable application.
    </Description>
    <Level>High</Level>
    <Match>eval\s?\(|setTimeout|setInterval</Match>
    <Result>Line: 20 - eval(opts.filter);</Result>
</report_mosca>
\end{minted}
\caption{\small Snippet of Mosca output classified with \textit{Score A}.}
\label{snippet:mosca-output-315}
\vspace{-0.2cm}
\end{listing}

\begin{listing}[t]
\begin{minted}[
    %xleftmargin=15pt,
    breaklines,
    breakbytoken,
    fontsize=\footnotesize]{text}
/src/lib/drivers/search/pouch.js-19- }
/src/lib/drivers/search/pouch.js:20: eval(opts.filter);
/src/lib/drivers/search/pouch.js-21- opts.filter = filter;
\end{minted}
\caption{\small Snippet of Graudit output classified with \textit{Score B}.}
\label{snippet:graudit-output-315}
\vspace{-0.27cm}
\end{listing}

\section{Effectiveness of the Tools (RQ3)}

In this section, we focus on our third research question (RQ3). To perform a quantitative and qualitative assessment of the selected tools, we begin by specifying the evaluation metrics and methodology we used to rank the tools. Then we present our findings, relying on the result of running the selected tools across all 957 advisories of our curated dataset.

\subsection{Evaluation Methodology}

\mypara{Tool evaluation metrics:} To evaluate the selected tools, we use two main metrics: \textit{true positive rate} (TPR) and \textit{precision} (P). The TPR represents the proportion of the total vulnerabilities that are correctly detected by a given tool, i.e. the true positives (TP): TPR=TP/|vulnerabilities|.
The TPR is useful to assess the raw detection rate of a tool without considering the influence of false positives (FP), i.e., its results that do not match the reported advisory. 
Precision represents the proportion of correctly classified positive cases: P=TP/(TP+FP). This metric is useful to assess if a tool produces too many false positives that can unnecessarily consume analysts' resources.

\mypara{Tool classification score:} To compute the evaluation metrics for a given tool, we need to analyze the output that it generates when applied to analyzing a specific vulnerability. Given that each tool outputs the vulnerability analysis results in its own specific, unstandardized format, we characterize a tool's output according to a common discrete classification score:

\begin{itemize}
\item \textbf{Score A}: The tool correctly detects and classifies the vulnerability reported in the advisory (\textit{true positive}).
\item \textbf{Score B}:  The tool shows a warning for the vulnerable code, but does not explicitly classify the finding as a vulnerability (\textit{true positive}).
\item \textbf{Score C}: The tool only shows results that do not match the vulnerability in the advisory report (\textit{false positives}).
\item \textbf{Score D}: The tool produces no output (\textit{false negative}).
\end{itemize}

We split the TP results according to two distinct classes: A means an \textit{explicit vulnerability notification}, and B a \textit{security warning notification}. The tools ranked with score A provide a richer output to the user and, thus, more information about the detected vulnerability.
As an example, consider the output of two tested tools, Mosca and Graudit, with regards to advisory 315 shown in Listings~\ref{snippet:mosca-output-315} and \ref{snippet:graudit-output-315}, respectively. Although both outputs flag the vulnerable \textit{eval} call reported by the review file of Listing~\ref{snippet:review}, Mosca's output clearly identifies a possible code injection, provides a description, a severity level, and the line of code containing the vulnerability. On the other hand, Graudit only shows the vulnerable line of code without explaining how or why it flags that particular snippet. For this reason, the output of Mosca is classified with \textit{Score A} while the output of Graudit is classified with \textit{Score B}.

This discrete classification is also important to account for tools that might flag for the vulnerability at a place that is only close to it (textually, or on the AST). Considering this, we require that tools must clearly identify the vulnerable statement for some vulnerabilities, e.g., code injections and others that can typically be pinpointed to a single statement, while for other vulnerability types multiple lines-of-code are acceptable. Two authors performed a cross-check of all tools' outputs to guarantee fairness of the tool classification in these cases.

\begin{table}[t]
\centering
\small
\resizebox{\columnwidth}{!}{
\begin{tabular}{lrrrrrr}
\hline
 \textbf{Tool} & \textbf{Min} & \textbf{Max} & \textbf{Mean} & \textbf{St. Dev.} & \textbf{Q-90} & \textbf{Total} \\ \hline
 ODGen         & 1.236        & 3653.043     & 148.757       & 385.629           & 370.969       & 110823.8       \\
 CodeQL        & 1.712        & 736.546      & 119.570       & 98.755            & 177.550       & 28696.8        \\
 NodeJsScan    & 20.023       & 984.453      & 99.562        & 121.246           & 230.216       & 23795.4        \\
 ESLint SSC    & 0.592        & 3556.665     & 29.871        & 237.799           & 25.465        & 7139.1         \\
 Graudit       & 0.042        & 14.632       & 0.550         & 1.624             & 0.704         & 131.3          \\
 InsiderSec    & 0.000        & 243.000      & 5.749         & 25.186            & 7.000         & 1374.0         \\
 MS DevSkim    & 0.276        & 186.338      & 6.393         & 23.262            & 8.044         & 1527.9         \\
 Drek          & 0.300        & 6.649        & 1.022         & 0.865             & 1.949         & 244.3          \\
 Mosca         & 0.005        & 245.498      & 7.408         & 25.166            & 12.216        & 1770.5         \\

\hline
\end{tabular}
}
\caption{\small Summary statistics of the analysis times (in seconds) taken by the tested tools across all 957 reviewed advisories.}
\label{table:analysis-times}
\vspace{-0.3cm}
\end{table}

\subsection{Analysis Performance}

We gauge analysis performance by measuring tools' execution time for all 957 advisories on a machine with an Intel Xeon E3-1220 v3 @ 3.10GHz processor and 32GB of memory.

\begin{table*}[t]
\centering
\footnotesize
\tabcolsep=0.055cm
\resizebox{\textwidth}{!}{
\begin{tabular}{l|cc|cc|cc|cc|cc|cc|cc|cc|cc}
\hline
\multirow{2}{*}{\textbf{Scope}} & 
\multicolumn{2}{c}{\textbf{ODGen}} & 
\multicolumn{2}{c}{\textbf{CodeQL}} & 
\multicolumn{2}{c}{\textbf{NodeJsScan}} & 
\multicolumn{2}{c}{\textbf{ESLint SSC}} &
\multicolumn{2}{c}{\textbf{Graudit}} &
\multicolumn{2}{c}{\textbf{InsiderSec}} &
\multicolumn{2}{c}{\textbf{MS DevSkim}} &
\multicolumn{2}{c}{\textbf{Drek}} &
\multicolumn{2}{c}{\textbf{Mosca}} \\
& TP (\%)                     & FP (P\%)                    & TP (\%)                     & FP (P\%)                    & TP (\%)                     & FP (P\%)                    & TP (\%)                     & FP (P\%)                 & TP (\%)                     & FP (P\%)                 & TP (\%) & FP (P\%) & TP (\%)                    & FP (P\%)               & TP (\%)                     & FP (P\%)                & TP (\%)                     & FP (P\%)               \\\hline
 \multirow{2}{*}{\makecell{CWE-22}}    & \cellcolor{yellow!35}70     & \cellcolor{yellow!35}136    & \cellcolor{green!30}104     & \cellcolor{yellow!35}416    & \cellcolor{yellow!35}56     & \cellcolor{yellow!35}257    & \cellcolor{green!30}110     & \cellcolor{red!15}25467  & \cellcolor{green!30}122     & \cellcolor{red!5}3101    & 2       & 401      & 0                          & 368                    & 0                           & 1057                    & 0                           & 241                    \\
                                       & \cellcolor{yellow!35}(47.9) & \cellcolor{yellow!35}(34.0) & \cellcolor{green!30}(71.2)  & \cellcolor{yellow!35}(20.0) & \cellcolor{yellow!35}(38.4) & \cellcolor{yellow!35}(17.9) & \cellcolor{green!30}(75.3)  & \cellcolor{red!15}(0.4)  & \cellcolor{green!30}(83.6)  & \cellcolor{red!5}(3.8)   & (1.4)   & (0.5)    & (0.0)                      & (0.0)                  & (0.0)                       & (0.0)                   & (0.0)                       & (0.0)                  \\\arrayrulecolor{gray!25}\hline
 \multirow{2}{*}{\makecell{CWE-79}}    & 1                           & 33                          & \cellcolor{yellow!35}26     & \cellcolor{red!5}843        & 6                           & 924                         & \cellcolor{yellow!35}29     & \cellcolor{red!15}123477 & 11                          & 23634                    & 0       & 28       & 0                          & 3990                   & 0                           & 6353                    & 1                           & 1664                   \\
                                       & (1.0)                       & (2.9)                       & \cellcolor{yellow!35}(26.3) & \cellcolor{red!5}(3.0)      & (6.1)                       & (0.6)                       & \cellcolor{yellow!35}(29.3) & \cellcolor{red!15}(0.0)  & (11.1)                      & (0.0)                    & (0.0)   & (0.0)    & (0.0)                      & (0.0)                  & (0.0)                       & (0.0)                   & (1.0)                       & (0.1)                  \\\arrayrulecolor{gray!25}\hline
 \multirow{2}{*}{\makecell{CWE-400}}   & 4                           & 40                          & 13                          & 212                         & 2                           & 374                         & \cellcolor{yellow!35}39     & \cellcolor{red!15}21936  & 4                           & 2795                     & 0       & 22       & 0                          & 1026                   & 0                           & 74                      & 1                           & 271                    \\
                                       & (4.5)                       & (9.1)                       & (14.6)                      & (5.8)                       & (2.2)                       & (0.5)                       & \cellcolor{yellow!35}(43.8) & \cellcolor{red!15}(0.2)  & (4.5)                       & (0.1)                    & (0.0)   & (0.0)    & (0.0)                      & (0.0)                  & (0.0)                       & (0.0)                   & (1.1)                       & (0.4)                  \\\arrayrulecolor{gray!25}\hline
 \multirow{2}{*}{\makecell{CWE-78}}    & \cellcolor{yellow!35}22     & \cellcolor{yellow!35}40     & \cellcolor{green!30}43      & \cellcolor{red!5}416        & 2                           & 121                         & \cellcolor{yellow!35}29     & \cellcolor{red!15}7405   & 4                           & 1567                     & 0       & 15       & 0                          & 269                    & 3                           & 380                     & 3                           & 190                    \\
                                       & \cellcolor{yellow!35}(29.3) & \cellcolor{yellow!35}(35.5) & \cellcolor{green!30}(57.3)  & \cellcolor{red!5}(9.4)      & (2.7)                       & (1.6)                       & \cellcolor{yellow!35}(38.7) & \cellcolor{red!15}(0.4)  & (5.3)                       & (0.3)                    & (0.0)   & (0.0)    & (0.0)                      & (0.0)                  & (4.0)                       & (0.8)                   & (4.0)                       & (1.6)                  \\\arrayrulecolor{gray!25}\hline
 \multirow{2}{*}{\makecell{CWE-818}}   & 0                           & 11                          & \cellcolor{yellow!35}16     & \cellcolor{yellow!35}57     & 0                           & 97                          & 1                           & 6990                     & 3                           & 1431                     & 1       & 8        & \cellcolor{green!30}64     & \cellcolor{red!5}1093  & 0                           & 393                     & 0                           & 150                    \\
                                       & (0.0)                       & (0.0)                       & \cellcolor{yellow!35}(21.3) & \cellcolor{yellow!35}(21.9) & (0.0)                       & (0.0)                       & (1.3)                       & (0.0)                    & (4.0)                       & (0.2)                    & (1.3)   & (11.1)   & \cellcolor{green!30}(85.3) & \cellcolor{red!5}(5.5) & (0.0)                       & (0.0)                   & (0.0)                       & (0.0)                  \\\arrayrulecolor{gray!25}\hline
 \multirow{2}{*}{\makecell{CWE-471}}   & \cellcolor{yellow!35}11     & \cellcolor{yellow!35}19     & \cellcolor{yellow!35}13     & \cellcolor{yellow!35}54     & 0                           & 295                         & \cellcolor{green!30}38      & \cellcolor{red!15}7466   & 0                           & 2632                     & 0       & 0        & 0                          & 249                    & 0                           & 220                     & 0                           & 438                    \\
                                       & \cellcolor{yellow!35}(22.9) & \cellcolor{yellow!35}(36.7) & \cellcolor{yellow!35}(27.1) & \cellcolor{yellow!35}(19.4) & (0.0)                       & (0.0)                       & \cellcolor{green!30}(79.2)  & \cellcolor{red!15}(0.5)  & (0.0)                       & (0.0)                    & (0.0)   & (0.0)    & (0.0)                      & (0.0)                  & (0.0)                       & (0.0)                   & (0.0)                       & (0.0)                  \\\arrayrulecolor{gray!25}\hline
 \multirow{2}{*}{\makecell{CWE-20}}    & 5                           & 19                          & 4                           & 25                          & 0                           & 116                         & \cellcolor{yellow!35}19     & \cellcolor{red!15}7947   & 4                           & 911                      & 0       & 13       & 0                          & 181                    & 1                           & 26                      & 2                           & 294                    \\
                                       & (12.2)                      & (20.8)                      & (9.8)                       & (13.8)                      & (0.0)                       & (0.0)                       & \cellcolor{yellow!35}(46.3) & \cellcolor{red!15}(0.2)  & (9.8)                       & (0.4)                    & (0.0)   & (0.0)    & (0.0)                      & (0.0)                  & (2.4)                       & (3.7)                   & (4.9)                       & (0.7)                  \\\arrayrulecolor{gray!25}\hline
 \multirow{2}{*}{\makecell{CWE-1321}}  & 3                           & 12                          & 5                           & 78                          & 0                           & 92                          & \cellcolor{green!30}31      & \cellcolor{red!15}18130  & 0                           & 4468                     & 0       & 9        & 0                          & 0                      & 0                           & 301                     & 0                           & 465                    \\
                                       & (8.3)                       & (20.0)                      & (13.9)                      & (6.0)                       & (0.0)                       & (0.0)                       & \cellcolor{green!30}(86.1)  & \cellcolor{red!15}(0.2)  & (0.0)                       & (0.0)                    & (0.0)   & (0.0)    & (0.0)                      & (0.0)                  & (0.0)                       & (0.0)                   & (0.0)                       & (0.0)                  \\\arrayrulecolor{gray!25}\hline
 \multirow{2}{*}{\makecell{CWE-94}}    & 4                           & 18                          & \cellcolor{yellow!35}5      & \cellcolor{red!5}79         & 2                           & 159                         & \cellcolor{yellow!35}16     & \cellcolor{red!15}17930  & \cellcolor{yellow!35}10     & \cellcolor{red!15}7159   & 1       & 23       & 0                          & 358                    & \cellcolor{yellow!35}8      & \cellcolor{red!15}858   & \cellcolor{yellow!35}8      & \cellcolor{red!5}199   \\
                                       & (12.1)                      & (18.2)                      & \cellcolor{yellow!35}(15.2) & \cellcolor{red!5}(6.0)      & (6.1)                       & (1.2)                       & \cellcolor{yellow!35}(48.5) & \cellcolor{red!15}(0.1)  & \cellcolor{yellow!35}(30.3) & \cellcolor{red!15}(0.1)  & (3.0)   & (4.2)    & (0.0)                      & (0.0)                  & \cellcolor{yellow!35}(24.2) & \cellcolor{red!15}(0.9) & \cellcolor{yellow!35}(24.2) & \cellcolor{red!5}(3.9) \\\arrayrulecolor{gray!25}\hline
 \multirow{2}{*}{\makecell{CWE-77}}    & \cellcolor{yellow!35}8      & \cellcolor{yellow!35}11     & \cellcolor{yellow!35}11     & \cellcolor{red!5}90         & 1                           & 32                          & \cellcolor{yellow!35}9      & \cellcolor{red!15}5390   & 0                           & 892                      & 0       & 1        & 0                          & 1                      & 0                           & 52                      & 0                           & 97                     \\
                                       & \cellcolor{yellow!35}(34.8) & \cellcolor{yellow!35}(42.1) & \cellcolor{yellow!35}(47.8) & \cellcolor{red!5}(10.9)     & (4.3)                       & (3.0)                       & \cellcolor{yellow!35}(39.1) & \cellcolor{red!15}(0.2)  & (0.0)                       & (0.0)                    & (0.0)   & (0.0)    & (0.0)                      & (0.0)                  & (0.0)                       & (0.0)                   & (0.0)                       & (0.0)                  \\\arrayrulecolor{gray!25}\hline
 \multirow{2}{*}{\makecell{Other CWE}} & 26                          & 154                         & \cellcolor{yellow!35}60     & \cellcolor{red!5}1283       & 34                          & 2548                        & \cellcolor{yellow!35}76     & \cellcolor{red!15}147552 & \cellcolor{yellow!35}61     & \cellcolor{red!15}60895  & 3       & 104      & 17                         & 7930                   & 3                           & 9159                    & 10                          & 2868                   \\
                                       & (8.9)                       & (14.4)                      & \cellcolor{yellow!35}(20.5) & \cellcolor{red!5}(4.5)      & (11.6)                      & (1.3)                       & \cellcolor{yellow!35}(26.0) & \cellcolor{red!15}(0.1)  & \cellcolor{yellow!35}(20.9) & \cellcolor{red!15}(0.1)  & (1.0)   & (2.8)    & (5.8)                      & (0.2)                  & (1.0)                       & (0.0)                   & (3.4)                       & (0.3)                  \\\arrayrulecolor{black}\hline
 \multirow{2}{*}{\makecell{Dataset}}   & \cellcolor{yellow!35}154    & \cellcolor{yellow!35}493    & \cellcolor{yellow!35}300    & \cellcolor{red!5}3553       & 103                         & 5015                        & \cellcolor{yellow!35}397    & \cellcolor{red!15}389690 & \cellcolor{yellow!35}219    & \cellcolor{red!15}109485 & 7       & 624      & 81                         & 15465                  & 15                          & 18873                   & 25                          & 6877                   \\
                                       & \cellcolor{yellow!35}(16.1) & \cellcolor{yellow!35}(23.8) & \cellcolor{yellow!35}(31.3) & \cellcolor{red!5}(7.8)      & (10.8)                      & (2.0)                       & \cellcolor{yellow!35}(41.5) & \cellcolor{red!15}(0.1)  & \cellcolor{yellow!35}(22.9) & \cellcolor{red!15}(0.2)  & (0.7)   & (1.1)    & (8.5)                      & (0.5)                  & (1.6)                       & (0.1)                   & (2.6)                       & (0.4)                  \\
\hline

\end{tabular}
}
\caption{\small TP, TPR (\%), FP and Precision (P\%) for each tool by CWE. TPR highlights: green (TPR $\ge$ 50\%) or yellow (50\% > TPR $\ge$ 15\%). When TPR is highlighted we also highlight the FP and P columns: yellow (50\% > P $\ge$ 15\%), light red (15\% > P $\ge$ 2.5\%) and dark red (P < 2.5\%). The CWEs are: Path Traversal (CWE-22), Cross-site Scripting (CWE-79), Resource Exhaustion (CWE-400), Insufficient Transport Layer Protection (CWE-818), OS Command Injection (CWE-78), Modification of Assumed-Immutable Data (CWE-471), Improper Input Validation (CWE-20), Code Injection (CWE-94), Improper Neutralization of Special Elements used in a Command (CWE-77), and Improperly Controlled Modification of Object Prototype Attributes (CWE-1321).}
\label{table:raw-detection}
\vspace{-0.2cm}
\end{table*}

Table~\ref{table:analysis-times} shows several statistics of the execution times taken by each tool to analyze our dataset. When compared to all other tools, ODGen, CodeQL, NodeJsScan and ESLint SSC require considerably more time. To analyze all 957 packages, ODGen took over 30 hours (110k seconds), CodeQL took nearly 8 hours (27k seconds), NodeJsScan took nearly 7 hours (24k seconds), while ESLint SSC took nearly 2 hours (7k seconds). All other tools are considerably more efficient, taking at most 30 minutes to analyze all packages.

The mean execution time of ODGen, CodeQL and NodeJsScan is 148.8, 119.6 and 99.6 seconds, respectively.
ODGen, CodeQL and NodeJsScan tools are slower because their detection techniques involve modeling statically computed structures. These operations are more complex than performing keyword-based matching searches (see Section~\ref{sec:methodologies}). Depending on the size of the package and on the CI/CD pipeline restrictions, these tools may end up being exceedingly slow.

\begin{figure}[t]
    \centering
    \includegraphics[width=0.8\columnwidth]{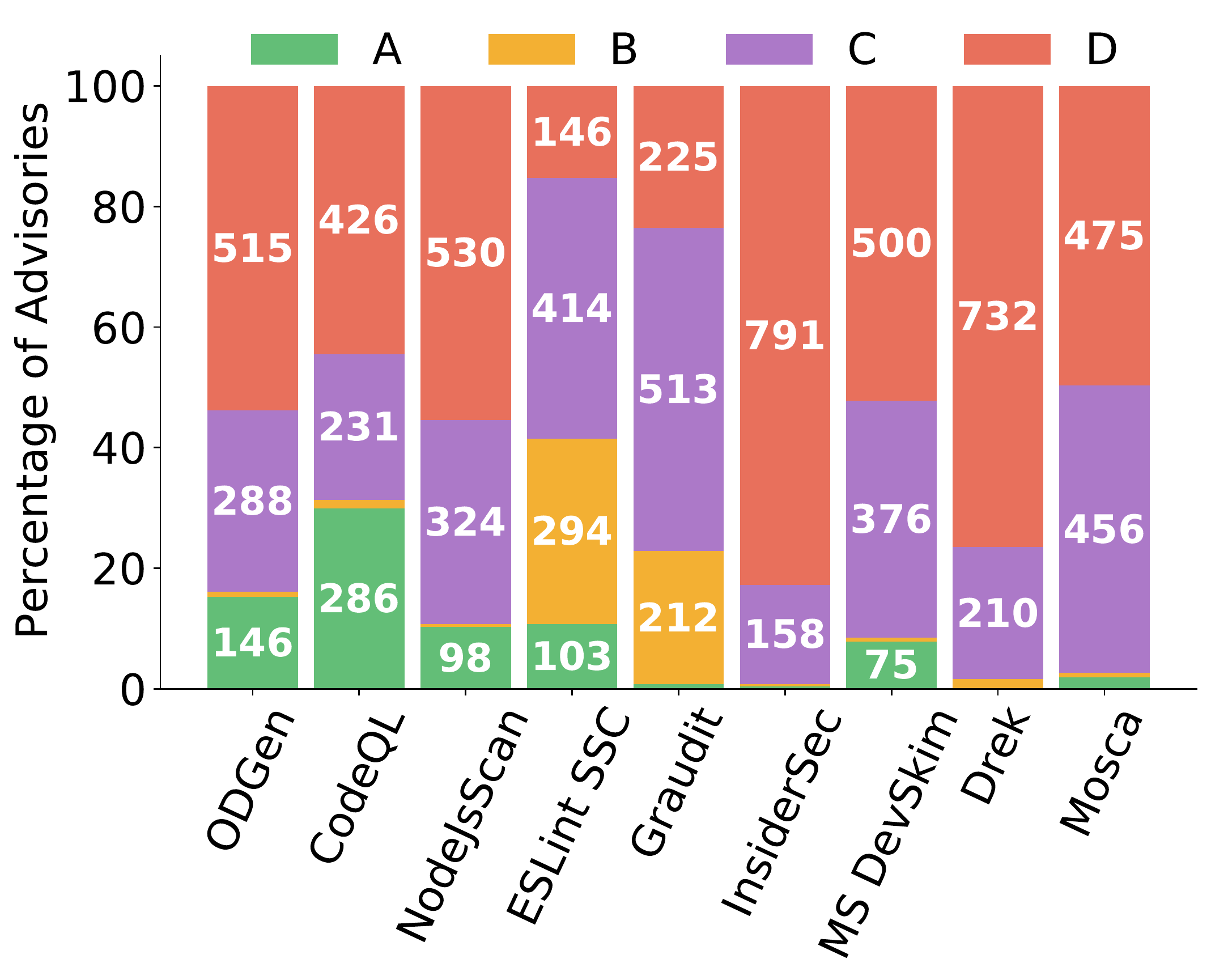}
    \vspace{-0.3cm}
    \caption{\small Score distribution for each tool.}
    \label{fig:tools-score-global}
    \vspace{-0.3cm}
\end{figure}%

\subsection{Results Across the Entire Dataset}

Figure~\ref{fig:tools-score-global} displays the score distribution for each tool across our entire dataset, and Table~\ref{table:raw-detection} shows the evaluation metrics for each tool. Globally, the tested tools perform rather poorly. We can draw the following main observations:

\mypara{1. Some tools have very low TPR:} Counting A and B scores as successful detections, we see that InsiderSec, Drek and Mosca only detect 7 (0.7\%), 15 (1.6\%) and 25 (2.6\%) vulnerabilities, respectively. Hence, these tools fail to detect most vulnerabilities of the dataset.

\mypara{2. The tools with best TPR have very low precision:} The tools that have higher TPR are: ODGen, Graudit, ESLint SSC, and CodeQL. Unfortunately, Graudit and ESLint SSC also have a considerable number of false positives, which tends to erode the confidence of application developers in vulnerability detection tools.
Graudit detects 219 vulnerabilities (22.9\%), but it also reports over 109k FPs, giving it an overall precision of just 0.2\%. A higher number of FPs is expected from a keyword-based tool like Graudit, as many of its string signatures often match non-vulnerable code snippets. 
ESLint SSC has the highest TPR (41.5\%). However, it is also the tool with the highest number of reported FPs (over 389k) and, consequently, the lowest precision (0.1\%). This is because ESLint SSC includes many rules from different ESLint plugins, some of which are simple matches (akin to keyword-based analysis) with greedy behaviour, leading to a higher number of FPs.

\begin{figure}[t]
    \centering
    \includegraphics[width=0.8\columnwidth]{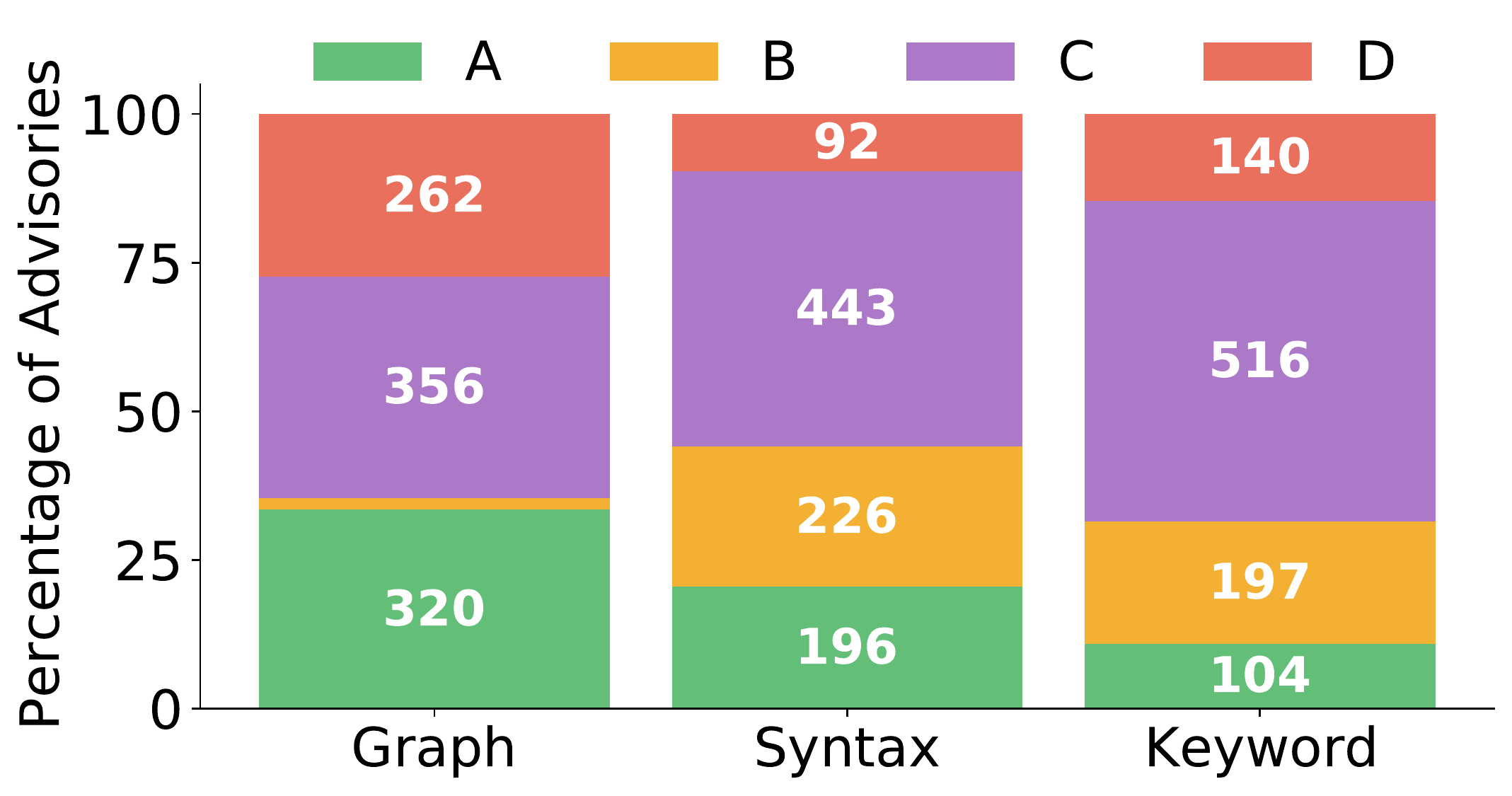}
    \caption{\small Score distribution for each detection technique.}
    \label{fig:tools-score-global-method}
    \vspace{-0.3cm}
\end{figure}

\mypara{3. Graph-based analysis has the best detection capability:} 
Figure~\ref{fig:tools-score-global-method} shows the scores according to a particular detection technique. These results show that graph-based analysis reports a significantly larger number of results with score A (explicit vulnerability notifications). Syntax and keyword-based analysis look fairly similar, with reasonable detection rates, but also a high number of reports containing only false positive results.

When considering the results of both tools in this category, i.e., ODGen and CodeQL, they strike a better balance between true positives and precision. CodeQL detects 300 vulnerabilities (31.3\%) and has a significantly higher precision (7.8\%), when compared to most other tools, while ODGen detects 154 vulnerabilities (16.1\%). This number is significantly lower than CodeQL's, but it represents a much higher precision (23.8\%) than any other tool tested. Although both these tools do not have the highest TPR, most of their detected vulnerabilities were classified with the A score, meaning that the reported information is richer and more meaningful to the user. Consequently, CodeQL and ODGen are the most balanced tools, achieving a reasonable detection rate (TPR) and less FPs, when compared to other tools with similar TPR. We also note that both these tools have the potential for being further improved by extending them with additional rules.

\mypara{4. Combining multiple tools increases TPR, but also lowers the overall precision:}
The combination of the two best tools (CodeQL and ESLint SSC) detects 508 vulnerabilities (53.1\%), albeit with only 0.12\% precision. If we add the third best tool (Graudit), we detect more vulnerabilities (551/57.6\%), but the precision further decreases to 0.11\%. Finally, combining both graph-based tools, CodeQL and ODGen, allows for the detection of 339 vulnerabilities (35.4\%) with a precision of 7.7\%. This shows that combining the best tools can increase the TPR, but at the cost of also increasing the number FPs, which limits the advantage of such an approach.

\subsection{Results Across Specific Vulnerability Types}

We now assess the performance of the tools when focusing on particular types of vulnerabilities. We concentrate on two main aspects: i) studying the types of vulnerabilities that the tools detect more frequently, and ii) analyzing which types of vulnerabilities can be detected simultaneously by several tools.

\mypara{1. Most frequently detected vulnerability types:} From the analysis of Table~\ref{table:raw-detection}, we  highlight seven CWEs that are detected most often regardless of the used tool: CWE-22, CWE-471, CWE-78, CWE-79, CWE-94, CWE-77, and CWE-1321. These are colored in yellow and green in Table~\ref{table:raw-detection}.

CWE-22 (path traversal) is the only type clearly detected by all four best performing tools (ODGen, ESLint SSC, Graudit, and CodeQL). This is because path traversal can be found statically by searching for well-known dangerous sinks in the Node.js API, e.g., the functions \textit{readFile}, \textit{writeFile} and \textit{createReadStream}. The difference in precision between these four tools lies in that ESLint SSC and Graudit simply match these function calls, while ODGen and CodeQL report only cases where the path is tainted by user input.

CWE-78 and CWE-77 (OS command injection), CWE-79 (cross-site scripting) and CWE-94 (code injection) correspond to classic injection vulnerabilities. Detecting these vulnerabilities depends on the sets of sinks considered by each tool. CodeQL detects more OS command injections, while ESLint SSC detects more code injections because each have more extensive rulesets for those particular vulnerabilities. Both tools detect about the same number of XSS vulnerabilities.

Both CWE-471 (Modification of Assumed-Immutable Data) and CWE-1321 (Improperly Controlled Modification of Object Prototype Attributes) are umbrella CWEs for several prototype tampering and prototype pollution vulnerabilities, for which both CodeQL and ESLint SSC have various rules. We expected ODGen to perform better at detecting prototype pollution vulnerabilities (CWE-471), as this is one of its central goals~\cite{odgen}. Although ODGen's results for this vulnerability (22.9\%) fall short of those by CodeQL (27.1\%) and ESLint SSC (79.2\%), ODGen does achieve a much higher precision (36.7\%) than CodeQL (19.4\%) and, especially, ESLint SSC (0.5\%).

\begin{figure}[t]
    \centering
    \includegraphics[width=\columnwidth]{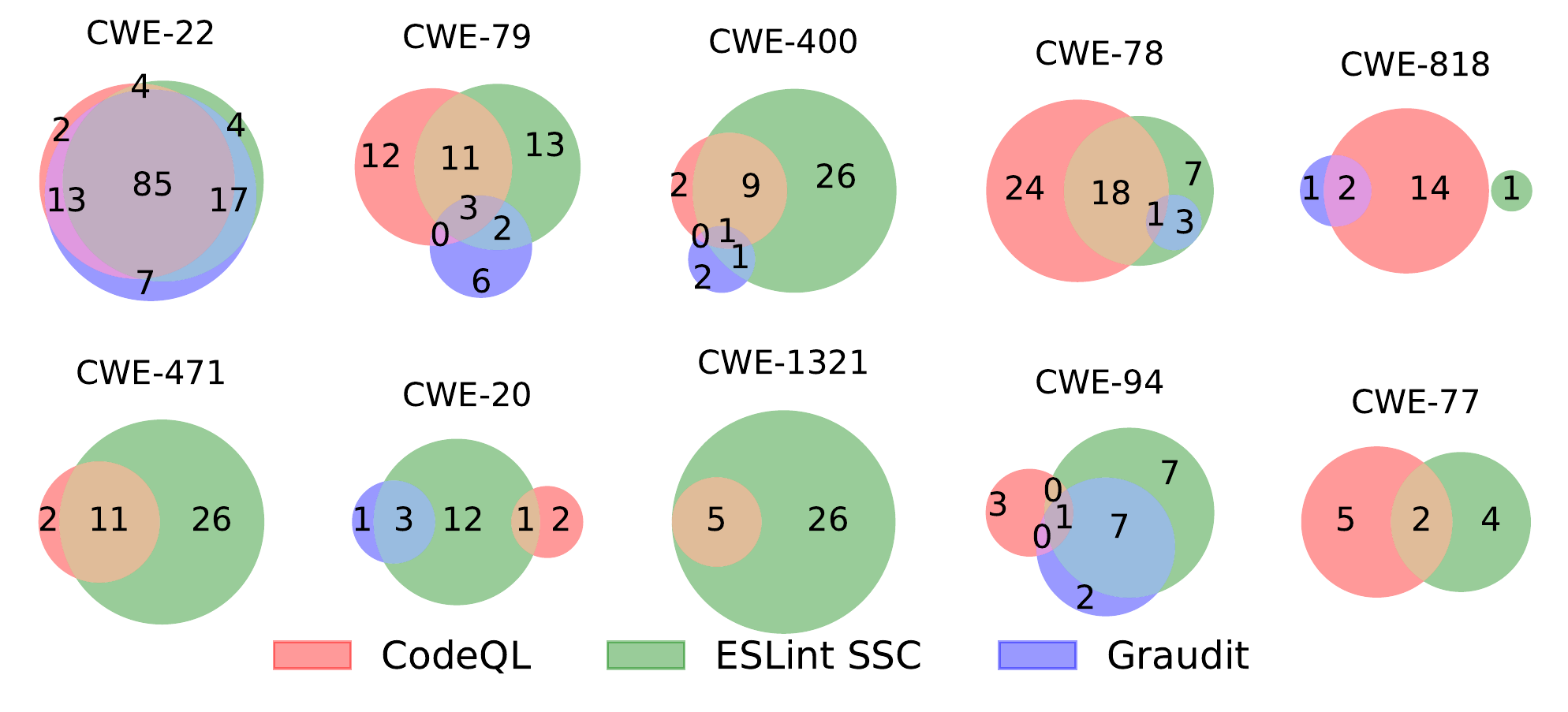}
    \vspace{-0.2cm}
    \caption{\small Intersections of TPs of the 3 best tools for top-10 CWEs.}
    \label{fig:cwe-venn}
    \vspace{-0.3cm}
\end{figure}

\mypara{2. Vulnerability types detected by the three best performing tools:} Figure~\ref{fig:cwe-venn} shows the intersections of TPs for the top-10 CWEs. We can see a substantial intersection for CWE-22, where all three tools detect the same 85 vulnerabilities. This happens because path traversals are easy to find statically using a limited set of known dangerous sinks from the Node.js API, which all tools share.
For CWE-79, both CodeQL and ESLint SSC detect about the same number of vulnerabilities, but only about half intersect with each other. This is due to differences in the rules regarding XSS sources and sinks. CWE-471 shows a significant intersection, but ESLint SSC detects several vulnerabilities that CodeQL misses. This is because ESLint SSC's rules have a wider range of sinks. Other CWEs have less intersections because their rulesets differ. For example, CodeQL is the only tool with specific rules to detect resource downloads over HTTP, hence the results for CWE-818.

\subsection{Results as a Function of Queries and Ruleset}

\begin{figure}[t]
    \centering
    \includegraphics[width=0.85\columnwidth]{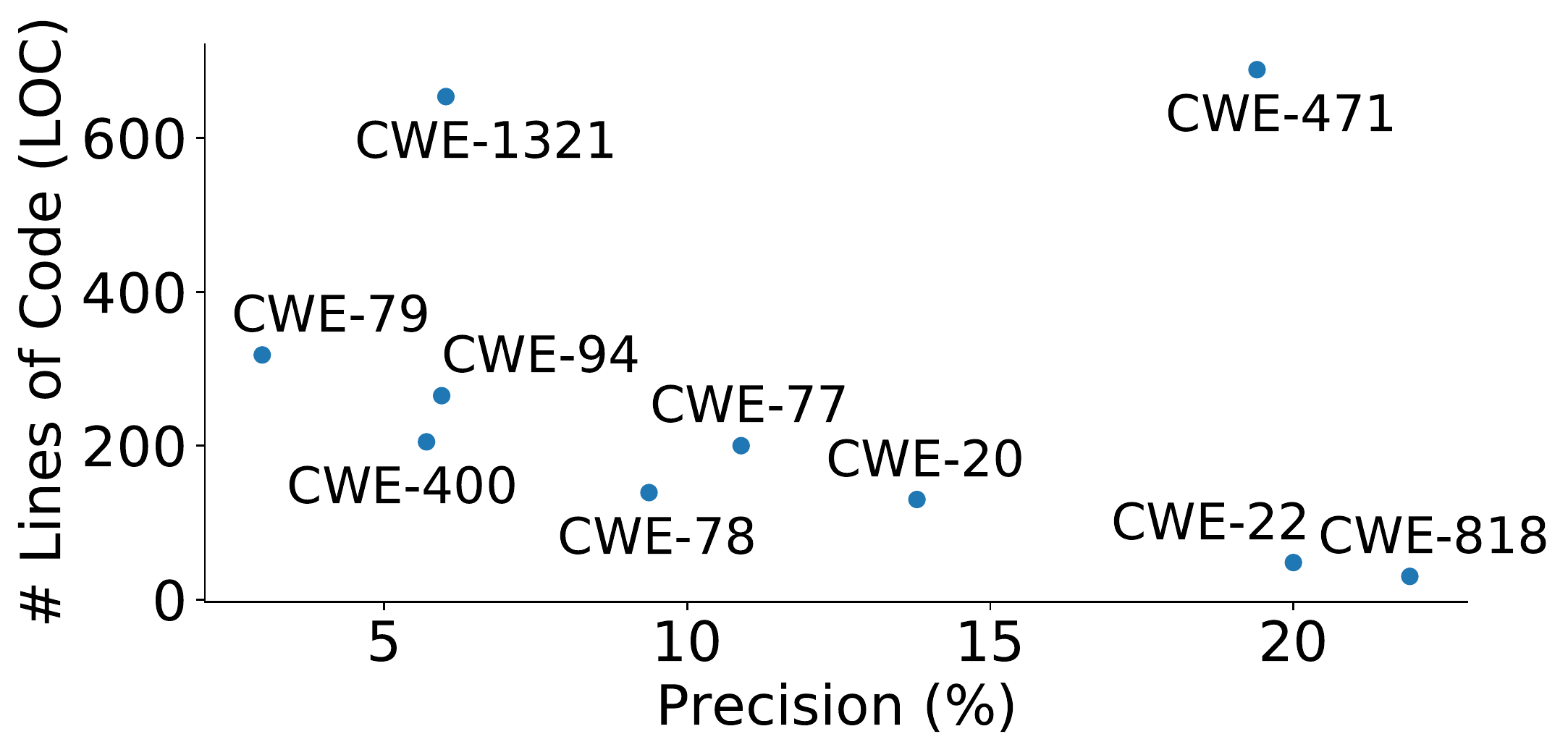}
    \vspace{-0.2cm}
    \caption{\small Correlation between Query LOC and Precision.}
    \label{fig:p-loc-correlation}
    \vspace{-0.3cm}
\end{figure}

To study the relationship between ruleset/queries and vulnerability detection results, we take CodeQL as an example and show, in Figure~\ref{fig:p-loc-correlation}, how the precision of this tool compares with the number of lines of code of the specific queries CodeQL uses to detect the vulnerabilities in the Top-10 CWE categories in the dataset. Although this cannot be taken as a general rule, we can see that, in most cases, the precision is higher for smaller queries. Taking into account each CWE, simpler vulnerabilities, like CWE-22, can be detected using smaller queries with higher precision, while more complex vulnerabilities to detect, like CWE-1312, are harder to detect even when using larger (more complex) queries.

It is then clear that vulnerability detection results can be influenced by the ruleset/queries executed by the tool. Using a small (specific) ruleset may improve the precision in detecting a specific vulnerability. However, in the context of a CI/CD pipeline, application developers do not know beforehand which specific ruleset to select to detect the (unknown) vulnerability. Consequently, it is reasonable to apply the most comprehensive ruleset available for the tool. This is the approach we used in this paper. For every tool, we selected the most comprehensive and complete ruleset, by either combining rules into a single tool execution or executing the tool multiple times using a different rule for every execution and combining the results. We also used the rules available off-the-shelf instead of developing or customizing rules. This allows us to reflect how developers will use these tools as most of them are not technically versed to improve the ruleset specified by the tool developers.

\section{Reasons for Missed Detection (RQ4)}
\label{sec:rq4}

In RQ4, we study why existing tools fail to detect certain vulnerabilities. Table~\ref{table:undetected-vulns} shows the number of undetected vulnerabilities grouped by CWE and their mapping to OWASP Top 10 Web Security Risks (2021)~\cite{owasp-mapping}. Of the 957 known vulnerabilities in the dataset, 324 vulnerabilities (33.9\%) were not detected by any of the selected tools. To understand the underlying reasons, we have manually analyzed a sample of undetected vulnerabilities. So far, we have identified the following five main tool limitations. In Table~\ref{table:limitations}, we map these limitations with some vulnerability categories (CWE) and provide specific examples of undetected vulnerabilities (CVE). 

\begin{table}[t]
\centering
\small
\tabcolsep=0.15cm
\resizebox{\columnwidth}{!}{
\begin{tabular}{llccc}
\hline
\textbf{CWE}    & \textbf{CWE Description}      & \textbf{OWASP}  & \textbf{Undetected}   & \textbf{\%} \\ \hline
CWE-79          & Cross-site Scripting          & \checkmark    & 50 / 99               & 50.5\% \\
CWE-400         & Resource Exhaustion           & -             & 44 / 89               & 49.4\% \\
CWE-78          & OS Command Injection          & \checkmark    & 20 / 75                & 26.7\% \\
CWE-20          & Improper Input Validation     & \checkmark    & 16 / 41                & 39.0\% \\
CWE-22          & Path Traversal                & \checkmark    & 12 / 146              & 8.2\% \\
CWE-94          & Code Injection                & \checkmark    & 10 / 33                & 30.3\% \\
CWE-818         & Insecure Transport Layer      & \checkmark    & 10 / 75               & 13.3\% \\
CWE-287         & Improper Authentication       & \checkmark    & 9 / 9                 & 100.0\% \\
CWE-471         & MAID                          & \checkmark    & 8 / 48                & 16.7\% \\
CWE-200         & Information Exposure          & \checkmark    & 8 / 14                & 57.1\% \\
Others          & -                             & -             & 137 / 286              & 47.9\% \\ \hline
Total           & -                             & -             & 324 / 957             & 33.9\% \\ \hline
\end{tabular}
}
\caption{\small Number of vulnerabilities undetected by any tool.}
\label{table:undetected-vulns}
\end{table}

    

\begin{table}[t]
\centering
\footnotesize
\tabcolsep=0.15cm
\begin{tabular}{ccll}
\hline
\textbf{Limitation} & \textbf{Advisory} & \textbf{CWE} & \textbf{Vulnerability} \\ \hline
\multirow{2}{*}{\makecell{L1}} &
    63 & CWE-730 & CVE-2015-9241\\ &
    567 & CWE-287 & CVE-2017-11429\\ \hline
\multirow{2}{*}{\makecell{L2}} & 
    165 & CWE-818 & CVE-2016-10583\\ &
    305 & CWE-22 & CVE-2016-1000249\\ \hline
\multirow{2}{*}{\makecell{L3}} & 
    26 & CWE-287 & CVE-2014-10067\\ &
    92 & CWE-200 & CVE-2016-10533\\\hline
\multirow{2}{*}{\makecell{L4}} &
    113 & CWE-89 & CVE-2016-10554\\ &
    43 & CWE-79 & CVE-2014-9772\\ \hline
\multirow{2}{*}{\makecell{L5}} &
    1469 & CWE-471 & CVE-2017-1000048\\ &
    313 & CWE-502 & CVE-2017-5954\\ \hline
\end{tabular}
\caption{\small Examples of undetected vulnerabilities by cause (Lx).}
\label{table:limitations}
\end{table}

\mypara{L1. Cross-package vulnerabilities}:
The selected tools come with pre-defined sets of manually written rules, typically focusing solely on popular APIs. We noticed that some undetected vulnerabilities exist in code that invokes functions of third-party packages that map directly to known dangerous code, e.g., wrappers to OS-level commands. These vulnerabilities could have been found by testing all package dependencies (can be thousands of other packages~\cite{zimmermann2019small}), or by using a more complete set of rules and queries (covering additional sources and sinks). However, the manual maintenance of such lists of sources and sinks is impractical as the Node.js ecosystem expands. Existing work~\cite{staicu2020extracting} tries to automatically extract taint specifications (sources and sinks) from JavaScript libraries, which partially solves the issue of incomplete rules, but requires the constant dynamic testing of every new \textit{npm} package.

\begin{listing}[t]
\begin{minted}[
    xleftmargin=15pt,
    breaklines,
    breakbytoken,
    linenos,
    fontsize=\footnotesize]{js}
// Snippet of ./gnuplot.js:
var run = require('comandante');

module.exports = function () {
    var plot = run('gnuplot', []);
    plot.print = function (data, options) {
        plot.write(data);
        // (...)
    };
    // (...)
}
\end{minted}
\caption{\small Command Injection (advisory 1440) - NPM and Github Advisories~\cite{advisory-1440, GHSA-cfwc-xjfp-44jg}.}
\label{snippet:advisory-1440}
\vspace{-0.2cm}
\end{listing}
For instance, Listing~\ref{snippet:advisory-1440} shows a code snippet of a command injection vulnerability for advisory 1440. This problem exists because user-controlled data reaches an \textit{exec} sink inside the third-party package \textit{comandante}, a package meant to ease the execution of OS-level commands. The tools fail to recognize this vulnerability because the usual command injection sinks are not directly present in the analyzed code, but are instead inside a third-party dependency that is not modelled by the vulnerability detection rules of each tool, i.e., they failed to include the \textit{write} function as a potential dangerous sink.

\mypara{L2. Limited analysis scope:} In addition to JavaScript code files, Node.js projects depend on several other components, such as configuration files, front-end template code, testing frameworks, etc. However, by analyzing only the JavaScript code in isolation, certain vulnerabilities can be missed.
As an example, \textit{npm} packages contain a \textit{package.json} file which may include bootstrap scripts. In several analyzed packages, these scripts are used to download resources over HTTP. As it turns out, using HTTP allows for man-in-the-middle attacks, where resources are replaced by malicious payloads. While some tools can detect insecure downloads if they are performed by the main JavaScript code (e.g., by searching for HTTP URLs), they cannot detect downloads issued from \textit{package.json}.

\begin{listing}[t]
\begin{minted}[
    breaklines,
    breakbytoken,
    fontsize=\footnotesize]{json}
{
    "scripts": {
        "preinstall":
        """wget http://s.qdcdn.com/17mon/17monipdb.zip &&
        unzip -p 17monipdb.zip 17monipdb.dat > 17monipdb.dat"""
    }
}
\end{minted}
\caption{\small Insecure Transport Layer in \textit{package.json} of \textit{ipip-coffee} package (advisory 279) - CVE-2016-10673.}
\label{snippet:advisory-279}
\vspace{-0.2cm}
\end{listing}
Listing~\ref{snippet:advisory-279} shows a snippet of the \textit{package.json} file for the \textit{ipip-coffee} package, in which an external resource is downloaded over HTTP. This allows for man-in-the-middle attacks that might compromise the server. In this particular example, this vulnerability can only be detected if the \textit{package.json} file is also considered when performing the vulnerability analysis.

\mypara{L3. Lack of contextual knowledge:} Packages may expose sensitive information, e.g., by logging plaintext passwords to a file. These vulnerabilities are application-specific and require contextual knowledge of which data is sensitive. The analyzed tools, however, are not designed to gain contextual knowledge and thus miss vulnerabilities that depend upon it, e.g., application-specific leaks. 
To help detect such vulnerabilities, a possible approach is to annotate application inputs, objects, or data flows with sensitivity levels, and check which system resources handle the annotated features during the execution.

\begin{listing}[t]
\begin{minted}[
    xleftmargin=15pt,
    breaklines,
    breakbytoken,
    linenos,
    fontsize=\scriptsize]{js}
// Snippet of ./lib/odbc.js:
if(exports.debug) {
    console.log("""%s odbc.js : pool[%s] :
        pool.close() - processing pools %s - connections: %s""",
        getElapsedTime(), self.index, key, connections.length);
}
\end{minted}
\caption{\small Credential Exposure (advisory 1185) - SNYK-JS-IBMDB-459762~\cite{SNYK-JS-IBMDB-459762}.}
\label{snippet:advisory-1185}
\vspace{-0.2cm}
\end{listing}

Listing~\ref{snippet:advisory-1185} shows an example of a Credential Exposure vulnerability, in which plaintext passwords are logged to the console. The code snippet itself seems benign until one becomes aware that the \textit{key} variable holds security-critical information. This contextual knowledge is needed to detect the vulnerability but is difficult to extract using automated tools.

\mypara{L4. Incorrect sanitization}: Application developers often use regular expressions to detect malicious inputs. However, regular expressions are complex, and developers usually do not test them thoroughly, allowing sanitization bypasses to occur. Sanitization errors are often hard to detect statically, as they require dynamically testing each regular expression ensuring that they generate semantically valid inputs that can both bypass the validation and effectively trigger the vulnerability. 

\begin{listing}[t]
\begin{minted}[
    xleftmargin=15pt,
    breaklines,
    breakbytoken,
    linenos,
    fontsize=\scriptsize]{js}
// Snippet of ./protect/lib/rules/xss.js
const xssSimple = new RegExp('((%3C)|<)((%2F)|/)*[a-z0-9%]+((%3E)|>)', 'i') 
const xssImgSrc = new RegExp('((%3C)|<)((%69)|i|(%49))((%6D)
    |m|(%4D))((%67)|g|(%47))[^\n]+((%3E)|>)', 'i')

function isXss(value) {
    return xssSimple.test(value) || xssImgSrc.test(value)
}
// Example attack payload:
// <input type="image" src onerror="alert('XSS')">
\end{minted}
\caption{\small XSS (advisory 1116) - CVE-2018-1000160.}
\label{snippet:advisory-1116}
\vspace{-0.2cm}
\end{listing}
For instance, Listing~\ref{snippet:advisory-1116} shows a code fragment containing two regular expressions that aim to prevent potential XSS vulnerabilities. However, these regular expressions are not entirely correct as there still exist some specially crafted inputs, such as the one shown in the comment of Listing~\ref{snippet:advisory-1116}, that can bypass this validation and launch an XSS attack.

\mypara{L5. Inability to cope with JavaScript dynamicity}:
Specific features of JavaScript can lead to vulnerabilities that are hard to detect by static analysis tools. For example, object-based inheritance, extensible objects, and dynamic typing are key features of JavaScript, which can lead to prototype pollution, authentication bypass, and business logic vulnerabilities.

\begin{listing}[t]
\begin{minted}[
    xleftmargin=15pt,
    breaklines,
    breakbytoken,
    linenos,
    fontsize=\scriptsize]{js}
// Snippet of ./lib/parse.js:
module.exports = function (str, opts) {
    var options = opts || {};
    var tempObj = typeof str === 'string' ? parseValues(str, options) : str;
    var obj = options.plainObjects ? Object.create(null) : {};

    var keys = Object.keys(tempObj);
    for (var i = 0; i < keys.length; ++i) {
        var key = keys[i];
        var newObj = parseKeys(key, tempObj[key], options);
        obj = Utils.merge(obj, newObj, options);
    }
    return Utils.compact(obj);
};
// Proof-of-Concept exploit code:
qs.parse("]=toString", { allowPrototypes: false }) // {toString = true} <== prototype overwritten
\end{minted}
\caption{\small Prototype Override (advisory 1469) - CVE-2017-1000048.}
\label{snippet:advisory-1469}
\vspace{-0.2cm}
\end{listing}
Listing~\ref{snippet:advisory-1469} shows a type of Prototype Pollution vulnerability present in the \textit{qs} package, which is a \textit{querystring} parsing library that allows developers to create objects within query strings. For example, the string \texttt{'foo[bar]=baz'} is converted to the object \texttt{\{foo:\{bar:'baz'\}\}}. Usually, this package protects against attacks that try to overwrite the existing prototype properties of an object. However, in this vulnerable version, the protection can be circumvented by prefixing the name of the parameter with character \texttt{[} or \texttt{]}, as shown in the proof-of-concept exploit code shown in Listing~\ref{snippet:advisory-1469}. Consequently, calling \texttt{toString()} on the object will throw an exception. This can subvert the application logic, potentially allowing attackers to work around security controls, modify data, and make the application unstable. The selected tools miss this example because they fail to model how objects change depending on the instructions applied to them, specially the object prototype.

\mypara{}From these limitations, we can extract actionable insights on the applicability of static code analysis tools for vulnerability detection in Node.js code. On one hand, these tools can potentially overcome limitations L1 and L2 by both employing improved strategies for maintaining taint specifications, and by considering all the appropriate analysis scopes for Node.js code. On the other hand, every static analysis tool will struggle to overcome limitations L3, L4, and L5, because they fail to capture behavioral and contextual information that is only available at runtime when the package is executed with appropriate, and application-specific, test inputs. To this end, it seems that the approaches employed by current static vulnerability detection tools can mainly be used successfully to detect classic injection-style vulnerabilities even if all the tools tested in this paper cannot do so with reasonable precision.

\section{Threats to validity}

\noindent \textbf{1.} Even though our dataset is composed of real known-vulnerable \textit{npm} packages, there may be an implicit bias towards vulnerabilities that are easier to analyze and more common across different programming languages (i.e., not specific to JavaScript code). 
Thus, since our curated dataset may not be fully representative of all vulnerabilities in Node.js applications, a tool that can detect all the vulnerabilities of our dataset may still miss other unreported ones.

\noindent \textbf{2.} We may have missed some relevant tool, failed to evaluate an analyzer that excels above all tested tools in our study, or overlooked third-party detection rules that produce better results. To reduce this risk, we will promote the reproducibility of our evaluation by providing both the source code of \mysys and our curated dataset.

\noindent \textbf{3.} Both the labeling of vulnerable packages and identification of their vulnerable code snippets were performed manually. Given the challenges of manual code inspection, these annotations could be mislabeled. To mitigate this risk, all vulnerabilities were analysed by at least two authors at separate times and we will make our dataset available for public scrutiny.

\noindent \textbf{4.} A potential concern is whether our study is susceptible to survivor bias. For instance, assuming hypothetically that all the packages that we analyze had already been analyzed using CodeQL during the code development phase, and that the vulnerabilities reported by CodeQL had been accordingly fixed by the developer prior to package release on \textit{npm}, then the number of vulnerabilities effectively detected by CodeQL could be higher than those reported in our study. This would misleadingly suggest that the quality of CodeQL is worse than what it is in reality. Note, however, that such a comprehensive characterization of each tool is beyond the scope of this work. In our study, we concentrate on evaluating tools' ability to detect, not all possible vulnerabilities, but only those that have been officially reported in \textit{npm} packages already in production.

\section{Related Work}

The literature covers many tools for detecting vulnerabilities in Web applications, including static~\cite{dahse2014static,backes2017efficient,alhuzali2016chainsaw}, dynamic~\cite{kieyzun:icse:2009,mcallister:raid:2008}, and hybrid analysis tools~\cite{balzarotti2007multi,felmetsger2010toward,pellegrino2017deemon,alhuzali2018navex}, often combining different types of program analysis techniques, such as fuzzing (e.g.~\cite{kieyzun:icse:2009,mcallister:raid:2008}), control-flow and data-flow analysis (e.g.~\cite{backes2017efficient, balzarotti2007multi, pellegrino2017deemon, yamaguchi2014modeling}), and symbolic execution (e.g.~\cite{ alhuzali2018navex, alhuzali2016chainsaw,felmetsger2010toward}). The great majority of these tools is, however, aimed at PHP-based Web applications, with considerably fewer tools targeting JavaScript applications. Most of the existing tools for JavaScript are aimed at client-side JavaScript code and its specific vulnerabilities: for instance, DOM-based XSS~\cite{likies:ccs:2013,melicher2018riding}, unrestricted inclusion of third-party cross-origin scripts~\cite{musch:asiaccs:2019}, and potentially malicious flows via client-side persistent storage~\cite{steffens2019don}.

\mypara{Graph-based vulnerability scanners:} State-of-the-art static vulnerability analysis techniques often work by first computing a static model describing the dynamic behaviour of the application to be analyzed. Most notably, code property graphs (CPGs)~\cite{yamaguchi2014modeling} were proposed as a compact representation of an application's behaviour. With CPGs, one can encode specific vulnerability types as simple graph traversals, which can, in turn, be expressed using graph query languages and then executed on top of off-the-shelf graph databases (e.g. Neo4J~\cite{neo4j}). Code property graphs have successfully been applied to find SQL injection, XSS, and CSRF vulnerabilities in PHP applications \cite{backes2017efficient,pellegrino2017deemon}. Furthermore, they are at the core of CodeQL~\cite{codeql}. For JavaScript, code property graphs were employed by JAW~\cite{jaw} and ODGen~\cite{odgen}, for client-side and server-side JavaScript respectively. In our work, we have extensively evaluated CodeQL and ODGen as representative state-of-the-art, graph-based vulnerability scanners.

\mypara{Vulnerability studies \& analyzers for Node.js applications:} Unlike client-side JavaScript applications, which run in the browser, Node.js application code is not sandboxed. Recent empirical studies~\cite{abdalkareem2017dev,zimmermann2019small} have shown that, contrary to popular belief, \textit{npm} applications are often poorly maintained and tested, with a significant percentage (up to 40\%) of all packages depending on code with at least one publicly known vulnerability. Furthermore, after reviewing more than 200K \textit{npm} applications, 
Staicu et al.~\cite{staicu2018synode} concludes that 20\% of the analyzed applications either directly or indirectly make use of an injection~API. Despite this security-critical situation, there is only a small number of  research tools for detecting vulnerabilities in Node.js applications and their underlying infrastructure, most of which based on dynamic code analysis. For instance, Synode~\cite{staicu2018synode} aims to prevent injection attacks in Node.js applications, and NodeSec~\cite{gong:thesis:2018} aims to detect vulnerabilities in Node.js applications. The authors of~\cite{staicu2018freezing} and~\cite{davis:fse:2018} design specific dynamic analysis for finding regular expression denial of service (ReDoS) vulnerabilities. The authors of~\cite{xiao2021abusing} also apply dynamic analysis and symbolic execution to detect attacks that leverage hidden properties in client- and server-side JavaScript. There are also academic works that employ static analysis techniques for detecting vulnerabilities in Node.js, but most focus on detecting prototype pollution vulnerabilities~\cite{li2021detecting,kim2022dapp}. ODGen~\cite{odgen} is the only purely static code analysis tool developed by the academia that aims to detect several types of vulnerabilities in Node.js.

\mypara{Empirical studies of vulnerability analyzers:} 
Several empirical studies aim at characterizing the efficacy of existing white-box vulnerability detection tools (e.g.~\cite{durieux2020empirical, melicher2018riding,nunes:computing:2019}). Durieux et al.~\cite{durieux2020empirical} evaluated 9 automated analysis tools for Ethereum Smart Contracts. The authors created a curated dataset consisting of 69 annotated vulnerable smart contracts, as well as a \emph{raw} dataset consisting of 47,518 smart contracts. They report that only 42\% of the vulnerabilities on the annotated dataset were detected, with the highest ranking tool having an accuracy of 21\%. Melicher et al.~\cite{melicher2018riding} evaluated 3 automated static analysis tools for detecting DOM-based XSS in client-side JavaScript code (Esflow~\cite{esflow}, ScanJS~\cite{mozilla-scanjs}, and Burp Suite Pro~\cite{burp}). They created a dataset with 3219 confirmed vulnerabilities. However, many security flaws in server-side code for Node.js do not exist on the client-side (e.g., SQL injections), and vice-versa. As such, the dataset from~\cite{melicher2018riding} is not representative enough of server-side vulnerabilities. Finally, Nunes et al.~\cite{nunes:computing:2019} evaluate five free static analysis tools for detecting SQL injection and XSS vulnerabilities in PHP web applications using a dataset comprising 134 WordPress plugins. In contrast to the studies referenced above, our paper presents the first  empirical study targeting fully automated vulnerability detection tools for \textit{npm} packages. 
Our study comes with a comprehensive manually-annotated dataset based on confirmed real-world vulnerabilities.

\section{Conclusions}

This paper presented an empirical study of static analysis tools for detecting vulnerabilities in Node.js packages. To conduct this study, we built \mysys, an automated analysis framework, using which we created the largest known curated dataset of Node.js packages with well-characterized security vulnerabilities. Currently, our curated dataset includes 745 reviews that accurately identify the exact location of known vulnerabilities inside affected \textit{npm} packages. We found that the nine evaluated tools fail to detect many vulnerabilities and exhibit high false positive rates. Additionally, we show that many important vulnerabilities appearing in the OWASP Top-10 are not detected by any evaluated tool or even when using the combination of all tools.

We believe that our curated dataset will substantially contribute to enabling future research on automatic vulnerability detection tools for server-side JavaScript applications. To this end, we have made this dataset publicly available.

\bibliographystyle{IEEEtran}
\bibliography{references.bib}

\end{document}